\documentclass[reprint,superscriptaddress,amsmath,amsfonts,amssymb,aps,PRL,showkeys]{revtex4-2}
\usepackage[english]{babel}
\usepackage{hyperref}
\usepackage{graphicx}
\usepackage{mathrsfs}

\hypersetup{
  breaklinks = {true},
  citecolor = {blue},
  colorlinks = {true},
  linkcolor = {purple},
}

\usepackage{physics}
\usepackage{mathtools}
\usepackage[dvipsnames]{xcolor}
\usepackage{braket} 
\usepackage{upgreek}

\usepackage{ulem}

\usepackage{soul}
\usepackage[T1]{fontenc}

\begin{document}

\title{Emergent Orbital Hall Currents in Disordered and Amorphous Materials}

\
\author{Jorge Martínez Romeral}
\affiliation{Catalan Institute of Nanoscience and Nanotechnology (ICN2), CSIC and BIST, Campus UAB, Bellaterra, 08193 Barcelona, Spain}
\affiliation{Department of Physics, Campus UAB, Bellaterra, 08193 Barcelona, Spain}

\author{Thomas Galvani}
\affiliation{Catalan Institute of Nanoscience and Nanotechnology (ICN2), CSIC and BIST, Campus UAB, Bellaterra, 08193 Barcelona, Spain}
\affiliation{Apeiron Intelligence S.L., Startup Lab Barcelona Activa, Carrer Llacuna 162, oficina 214, 08018 Barcelona}

\author{Onurcan Kaya}
\affiliation{Catalan Institute of Nanoscience and Nanotechnology (ICN2), CSIC and BIST, Campus UAB, Bellaterra, 08193 Barcelona, Spain}

\author{Stephan Roche}
\affiliation{Catalan Institute of Nanoscience and Nanotechnology (ICN2), CSIC and BIST, Campus UAB, Bellaterra, 08193 Barcelona, Spain}
\affiliation{ICREA--Instituci\'o Catalana de Recerca i Estudis Avan\c{c}ats, 08010 Barcelona, Spain}

\author{Aron W. Cummings}
\email{aron.cummings@icn2.cat}
\affiliation{Catalan Institute of Nanoscience and Nanotechnology (ICN2), CSIC and BIST, Campus UAB, Bellaterra, 08193 Barcelona, Spain}

\author{Luis M. Canonico}
\email{luis.canonico@icn2.cat}
\affiliation{Catalan Institute of Nanoscience and Nanotechnology (ICN2), CSIC and BIST, Campus UAB, Bellaterra, 08193 Barcelona, Spain}

\date{\today}

\begin{abstract}
The role of disorder in orbital transport remains a central open question in orbitronics. Using a real-space formulation of itinerant orbital angular momentum, we investigate the real-time orbital Hall transport in single-orbital square-lattice and graphene-based systems. We demonstrate that both structural (polycrystalline and amorphous) and electrostatic disorder can generate sizable orbital Hall currents that are absent in their pristine counterparts. Remarkably, in amorphous graphene, these currents persist over extended timescales and across a broad energy window around charge neutrality, despite the strong localization of electronic states. Consistent with recent experimental evidence, our findings establish amorphous quantum materials as promising platforms for disorder-engineered orbitronic functionality.


\end{abstract}

\maketitle
\textit{Introduction} - Orbitronics, which seeks to control and transport angular momentum in solids, has emerged as a promising platform for doing so without relying on strong spin–orbit coupling \cite{go2021orbitronics,burgos2024orbital,ando2025orbitronics,cysne2025orbitronics}. Developments in this field have broadened the understanding of angular momentum transport by revealing an intricate interplay between orbital, spin, and lattice degrees of freedom. Interest in these phenomena increased significantly following the experimental observation of the orbital Hall effect (OHE) \cite{choi2023observation,Lyalin2023} and earlier theoretical predictions \cite{Bernevig2005,Go2018,bhowal2020intrinsic,canonico2020orbital}. Orbital effects are now being actively explored for magnetization control in nonvolatile memories \cite{GoOrbitalTorque2020,lee2021orbital,hayashi2023observation,gupta2025}, spin–orbital interconversion \cite{Santos2023,santos2024exploring,hayashi2024observation,abrao2025anomalous,costa2025dominance}, and THz emission driven by the inverse orbital Hall and orbital Rashba–Edelstein effects \cite{seifert2023,xu2024orbitronics,liu2024efficient}. Together, these developments establish orbital angular momentum (OAM) not only as a fundamental component of nonequilibrium transport phenomena but also as a promising platform for next-generation spin-orbitronic devices based on the manipulation of angular momentum flows.

Despite this rapid progress, a central question remains unresolved: \textit{what is the role of disorder in the electrical generation of orbital responses?} While theoretical studies focused primarily on intrinsic OHE, arising from orbital textures in reciprocal space and pristine crystalline systems \cite{Go2018,bhowal2020intrinsic,canonico2020orbital}, experimental samples inevitably contain disorder in the form of structural imperfections or electrostatic inhomogeneities. 
In this context, recent experiments of orbital pumping in polycrystalline systems suggest ballistic propagation of OAM \cite{seifert2023}, while nonlocal measurements in naturally oxidized Cu systems exhibit robust long-range orbital transport signatures with long diffusion lengths \cite{gao2025nonlocal}. Furthermore, amorphous Mo samples employed in spin–orbit torque devices display critical currents comparable to those of Pt- and Ir-based devices despite their substantially weaker spin–orbit coupling \cite{wang2024}. Finally, recent torque measurements in disordered SrRuO${}_3$-based devices \cite{peng2025unconventional} and MOKE measurements in amorphous Mn \cite{choi2026electron} report a surprising enhancement of orbital currents due to the disorder. 

The theoretical role of disorder in orbital transport remains also unsettled. While atom-centered approaches predict a strong suppression of the OHE by scattering events \cite{tang2024role,rang2024orbital}, theories accounting for itinerant orbital motion reveal substantial extrinsic contributions, potentially exceeding the intrinsic response \cite{Walz2014,veneri2025extrinsic,Joy2026Skew}. Real-space studies further indicate that orbital Hall responses can survive or even be enhanced by disorder \cite{Vidarte2026real,canonico2024orbital,Barbosa2026Extrinsic}. Yet whether this resilience extends to strongly localized electronic states and realistic structural disorder remains unknown. Addressing this question requires approaches that resolve orbital dynamics beyond crystalline momentum-space descriptions. Large-scale real-space and time-dependent quantum-transport methods \cite{seifert2023,gao2025nonlocal,wang2024,peng2025unconventional,choi2026electron} provide the necessary framework to explore this regime. 

Using these techniques, we investigate the OHE in systems with vanishing intrinsic orbital Hall response, which allows us to disentangle intrinsic and disorder-induced contributions. We consider square lattices and graphene subjected to electrostatic and Anderson disorder, as well as polycrystalline and amorphous graphene. Our results reveal that, independently of the system, a pronounced orbital Hall response emerges when disorder is introduced, despite the absence of an intrinsic OHE. We also observe a decoupling of electronic and orbital transport: although the electronic states rapidly localize, the orbital Hall conductivity remains sizable and robust over very long times. These findings are consistent with recent experimental observations suggesting that disorder is not detrimental to orbital transport and may even enhance it \cite{seifert2023,gao2025nonlocal,wang2024,peng2025unconventional,choi2026electron}.

\begin{figure}[tbh]
\includegraphics[width=1\columnwidth]{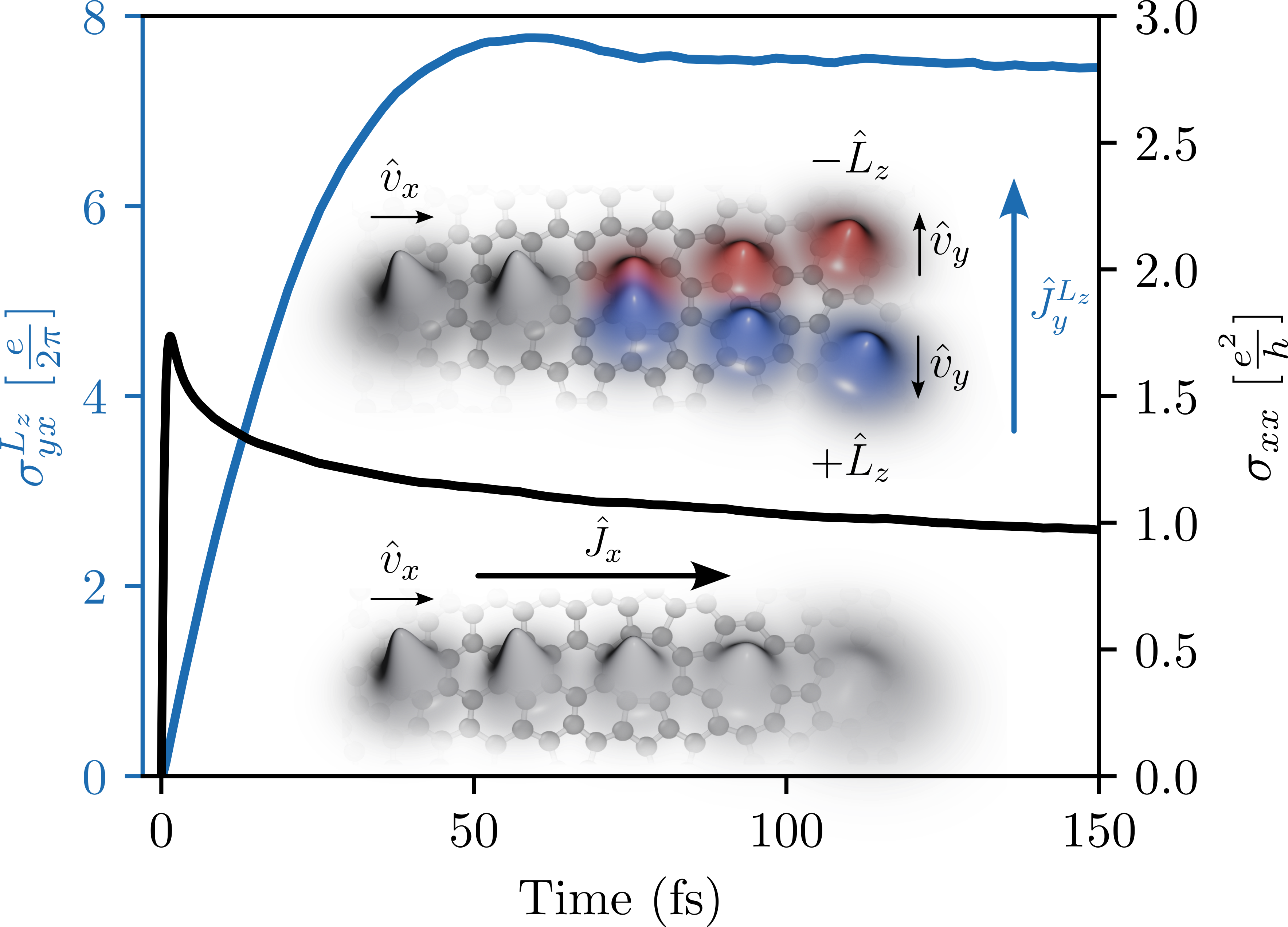}
\caption{\label{fig:scheme} Itinerant orbital Hall conductivity (blue) and longitudinal charge conductivity (black) for an $s$-orbital square lattice with Anderson disorder $W=2$ eV. In both simulations $\mu=0$ eV and $T=300$ K. Inset: schematic representation of the formation of the orbital Hall effect and the decrease of the longitudinal transport.}
\end{figure}

\begin{figure*}[tbh]
\includegraphics[width=1\textwidth]{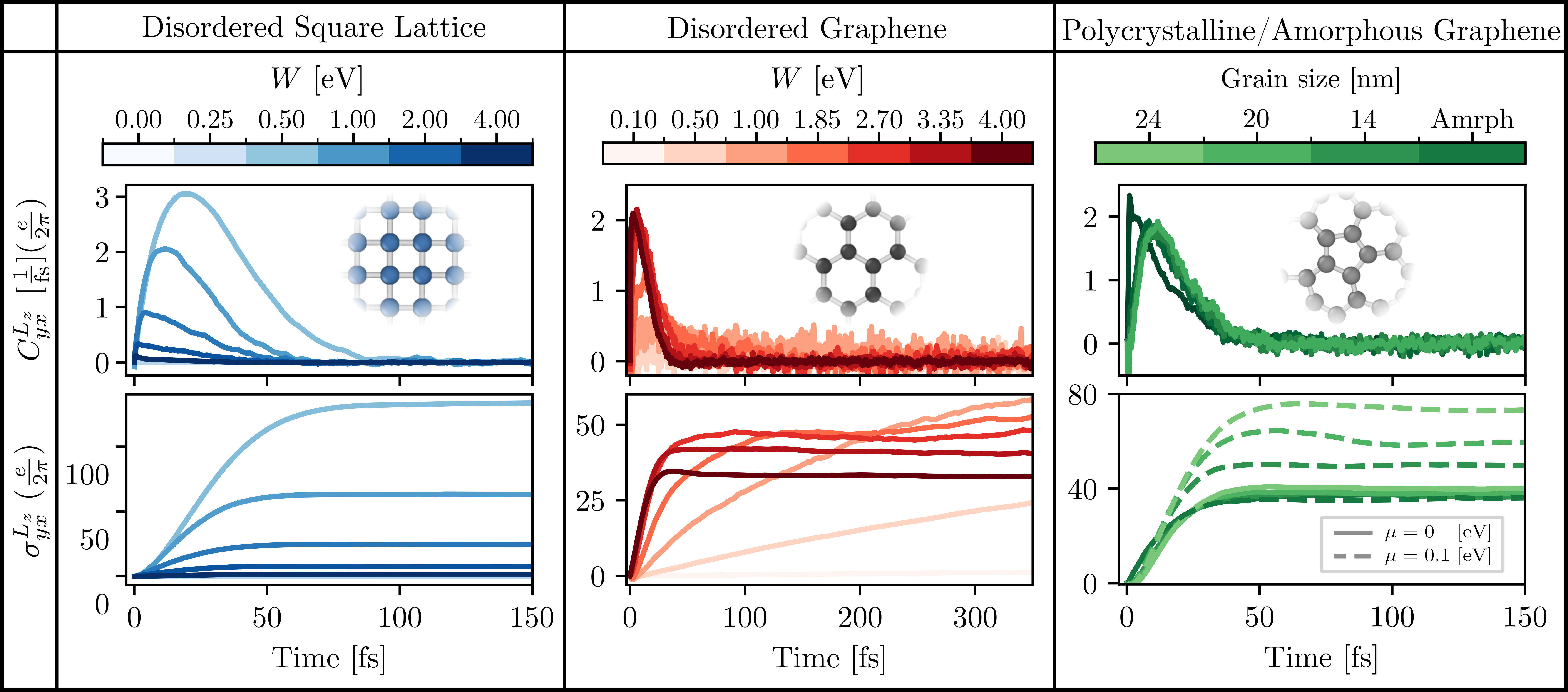}
\caption{\label{fig:square_graphene_amorph} Time evolution of the orbital Hall current correlation function (top) and conductivity (bottom) for an s-orbital square lattice (left) and graphene (center) with varying Anderson disorder strengths, and polycrystalline to fully amorphous graphene (right) with decreasing grain size. Solid (dashed) curves correspond to chemical potentials $\mu = 0$ $(0.1)$~eV in the right panels; $\mu=0$ elsewhere. All calculations use $T=300$ K, $M=1024$ Chebyshev moments, and $10^6$-$10^7$ orbitals.}
\end{figure*}

\textit{Methodology} - Our methodology centers around the time evolution of the electronic density matrix, $\hat{\rho}(t)$, in the presence of an external perturbation, by solving the quantum Liouville equation, $i\hbar\partial_t\hat{\rho}(t) = \left[\hat{H} (t),\hat{\rho}(t)\right]$ \cite{Aversa1995}. The Hamiltonian, $\hat{H}$, is equal to the sum of an equilibrium Hamiltonian, $\hat{H}_{\mathrm{eq}}$,  plus a perturbation term, $\hat{V}f(t)$, where $\hat{V}$ is the form of the perturbation and $f(t)$ describes its evolution in time. We are interested in computing the conductivity that emerges when slowly turning on an electric field $\mathbf{E}$ over a time $\tau_{\phi}$, by letting $\hat{V}f(t)=-e\mathbf{E}\cdot\mathbf{\hat{r}}\exp\left(t/\tau_{\phi}\right)$ where $e$ is the absolute value of the electron charge and $\mathbf{\hat{r}}$ is the position operator. Following Refs. \cite{zubarev1960,Romeral2025QuantumMetric,veiga2024unambiguous,santos2020landauer,diventra2008electrical}, we truncate at linear order in $\hat{V}$ and calculate the expectation value of a general current density in the $\alpha$ direction, $\hat{J}_{\alpha}$. Using $\braket{\hat{J}_{\alpha}}=\sigma_{\alpha\beta}E^{\beta}$ and after taking the adiabatic limit $\tau_{\phi}\to\infty$, the conductivity is given by 
\begin{equation}\label{eq:conductivity}
    \sigma_{\alpha\beta}=\dfrac{ie}{\hbar}\int_0^{\infty}dt'\mathrm{Tr}\left\{\left[\hat{r}_{\beta},\hat{\rho}_{\mathrm{eq}}\right]\hat{U}^{\dagger}(t',0)\hat{J}_{\alpha}\hat{U}(t',0)\right\},
\end{equation}
where $\hat{r}_{\beta}$ is the position operator along the direction $\beta$, $\hat{\rho}_{\mathrm{eq}}\equiv\hat{\rho}(\hat{H},\mu,T)=1/[1+\mathrm{exp}((\hat{H}-\mu)/k_{\mathrm{B}}T)]$ is the equilibrium density matrix described by the Fermi-Dirac distribution with chemical potential $\mu$ and temperature $T$, $\hat{U}(t',0)$ is the time evolution operator from $0$ to $t'$, and $\hbar$ is the reduced Planck constant (more details in the Supplementary material (SM) \cite{suppmaterial} and Ref. \cite{Romeral2026LightHall}). 
 
The expression \eqref{eq:conductivity} is equivalent to the Kubo-Bastin formula \cite{Streda1982Theory, Bastin1971Quantum} (see SM \cite{suppmaterial} for a comparison). Nevertheless, the present formalism has the key advantage of allowing us to calculate the time evolution of the electrical response via the time-dependent correlation function, $C_{\alpha\beta}(t)=\mathrm{Tr}\left\{\left[\hat{r}_{\beta},\hat{\rho}_{\mathrm{eq}}\right]\hat{J}_{\alpha}(t)\right\}$, where $\hat{J}_{\alpha}(t)=\hat{U}^{\dagger}(t,0)\hat{J}_{\alpha}\hat{U}(t,0)$,  which measures the correlation between the drive induced by an instantaneous pulse and the resulting current \cite{Kubo1957Stadistical, Bruus2004ManyBody,Reichl2016Modern}. Its integral leads to the time evolution of the conventional conductivity, Eq. \eqref{eq:conductivity}, enabling the identification of different transport regimes. Such responses can only be examined within time-dependent approaches \cite{PanhansOrtmann2021}, in contrast to the conventional spectral representation of the Kubo formula ~\cite{Streda1982Theory, Bastin1971Quantum, fan2021linear}.

We focus on the computation of the itinerant orbital Hall conductivity  $\sigma_{\alpha\beta}^{L_\gamma}$. The itinerant orbital current operator is $\hat{J}^{L_{\gamma}}_{\alpha}=\frac{1}{2}\{\hat{L}_{\gamma},\hat{v}_{\alpha}\}/\Omega$, where $\hat{L}_{\gamma}=\epsilon_{\gamma\nu\xi}\frac{e\hbar}{4\mu_B}(\hat{r}_{\nu}\hat{v}_{\xi}-\hat{r}_{\xi}\hat{v}_{\nu})$, $\hat{v}_{\xi}$ is the velocity operator along axis $\xi$, $\epsilon_{\gamma\nu\xi}$ is the Levi-Civita tensor, $\mu_B$ is the Bohr magneton and $\Omega$ is the area of the system \cite{Tonhauser2005,Bhowal2021}. Although $\sigma_{\alpha\beta}^{L_\gamma}$ derives generally from  atomic angular momentum contributions \cite{canonico2020orbital}, we consider spinless systems with single-orbital character, for which the atomic orbital moment vanishes. This leaves the itinerant contribution as the sole source of orbital Hall response.

To evaluate the trace in Eq.\ \eqref{eq:conductivity}, we use the stochastic approximation \cite{suppmaterial, Weisse2006Kernel, fan2021linear}. For functions of the Hamiltonian, such as $\hat{U}(t_1,t_0)$ and the Green's functions entering the $\hat{L}_\gamma$ operator \cite{canonico2024orbital}, instead of diagonalizing $\hat{H}$ we employ the kernel polynomial method (KPM) and expand them as a series of Chebyshev polynomials \cite{suppmaterial, Weisse2006Kernel, fan2021linear}. Terms including the position or velocity operator can be efficiently evaluated in periodic systems thanks to our real-space formulation \cite{Romeral2025QuantumMetric}. The operator $\hat{L}_{\gamma}$ is computed using a real-space representation that considers the off-diagonal elements of the position operator \cite{canonico2024orbital}. The combination of the Chebyshev polynomial expansion with the stochastic trace approximation makes the method $\mathcal{O}(N)$, where the simulation time scales linearly with the number of orbitals in the system, $N$. Here, we study systems with one to sixteen million orbitals, corresponding to sample sizes up to 0.7 $\upmu$m $\times$ 0.7 $\upmu$m.

We consider three systems: a square lattice with one $s$-orbital per site, graphene, and polycrystalline/amorphous graphene, the latter two with one $p_z$-orbital per site~\cite{FoaTorresRocheChalier2014}. The square lattice and graphene are described by a nearest-neighbor tight-binding (TB) Hamiltonian, $\hat{H} = \sum_{\left<ij\right>}  \gamma_{ij} c_i^{\dagger} c_j +\sum_i\epsilon_ic_i^{\dagger}c_i$, where $\gamma_{ij}$ is the nearest-neighbor hopping. For the square lattice and graphene we let $\gamma_{ij} = -1$~eV and $-2.7$~eV, respectively. The onsite term is used to model Anderson disorder \cite{Anderson1958}, with $\epsilon_i$ chosen randomly in the uniform distribution $[-W,W]$.
The polycrystalline and amorphous samples are described by a TB Hamiltonian that includes renormalization of the onsite energies and hoppings of atoms belonging to non-hexagonal rings \cite{cummings2019universal,suppmaterial}. The polycrystalline samples were created with molecular dynamics using a nucleation and annealing approach meant to mimic chemical vapor deposition (CVD) ~\cite{Kotakoski2012Mechanical}, while the amorphous sample was built using a melt-quench protocol \cite{Lindsay2010Optimized,suppmaterial}.

\textit{OHE in the disordered square lattice} -  We begin by studying the orbital Hall response in the simplest 2D system, the square lattice.
An example of the dynamics is shown in Fig.~\ref{fig:scheme}, for an Anderson disorder strength of $W=2$~eV.
The black line shows the time-dependent longitudinal conductivity, $\sigma_{xx}$, calculated from the cumulative integral of Eq.~\eqref{eq:conductivity} with $\hat{J}_\alpha = -e \hat{v}_\alpha / \Omega$.
For this disorder strength the system quickly enters the localization regime, characterized by an exponential decreasing $\sigma_{xx}$ with time. 
The blue line shows $\sigma_{yx}^{L_z}(t)$. Surprisingly, disorder gives rise to a finite orbital Hall conductivity, even though the clean square lattice exhibits no intrinsic orbital Hall response. Moreover, unlike charge transport, which progressively vanishes with time due to localization, the orbital Hall conductivity remains finite and eventually saturates, although its absolute value decays with disorder strength as discussed below.

The growth of $\sigma_{yx}^{L_z}$ at short times can actually be qualitatively understood by considering a wave packet driven by an electric field along the $x$ direction. Before scattering, the wave packet has a finite longitudinal velocity, $\braket{v_x}\neq0$, and no net transverse motion, $\braket{v_y}=0$. When the wave packet encounters an impurity, scattering modifies its distribution in momentum and real space, generating local transverse velocity and orbital angular momentum. Symmetry can enforce that these local contributions cancel globally, such that $\braket{v_y}=0$ and $\braket{L_z}=0$. However, their product, in the presence of an electric field along $x$, is not required to vanish upon averaging, giving rise to finite orbital Hall transport, $\sigma_{yx}^{L_z}  \propto \braket{\hat{v}_x \{\hat{L}_z , \hat{v}_y \}}  \neq 0$, which can build up over time with successive scattering events. This qualitative picture of the disorder-induced generation of correlated transverse motion and orbital angular momentum is illustrated in the inset of Fig. \ref{fig:scheme}.

The bridge between this wave-packet picture and the Kubo formula of Eq.~\eqref{eq:conductivity}, which accounts for the entire electronic ground state, can be made explicit by examining the orbital-charge current correlation function, given by the integrand of Eq. \eqref{eq:conductivity}. This correlation function, plotted in the first row of Fig. \ref{fig:square_graphene_amorph}, describes the correlation between the velocity of the ground state in the $x$ direction and the time-dependent transverse orbital current, $J_{y}^{L_z}(t)$, such that $C_{yx}^{L_z}(t)\propto\braket{\hat{v}_x(0)\hat{J}_{y}^{L_z}(t)}+\mathcal{O}(\hat{v}_x^2)$, as at first order $\left[\hat{r}_{x},\hat{\rho}_{\mathrm{eq}}\right]\propto\hat{v}_x+\mathcal{O}(\hat{v}_x^2)$. In the clean limit, since there is no scattering and no generated correlation between transverse velocity and angular momentum, the orbital current is $J_{y}^{L_z}(t)=0$ for all times. In contrast, disorder-induced scattering drives the correlation to grow exponentially toward a maximum over a rise time $\tau_r$, followed by an exponential decay over a time scale $\tau_d$ due to the loss of correlation between the electric field perturbation and the orbital Hall response, as shown in Fig. \ref{fig:square_graphene_amorph}. The source of this decay depends on the transport regime, and will be revisited below when we discuss the graphene case. Nevertheless, $C_{yx}^{L_z}$ never becomes negative; as there is no anticorrelation, the conductivity never starts to decrease, in contrast to $\sigma_{xx}(t)$ which is driven to zero by localization effects.

\begin{figure}[tbh]
\includegraphics[width=1\columnwidth]{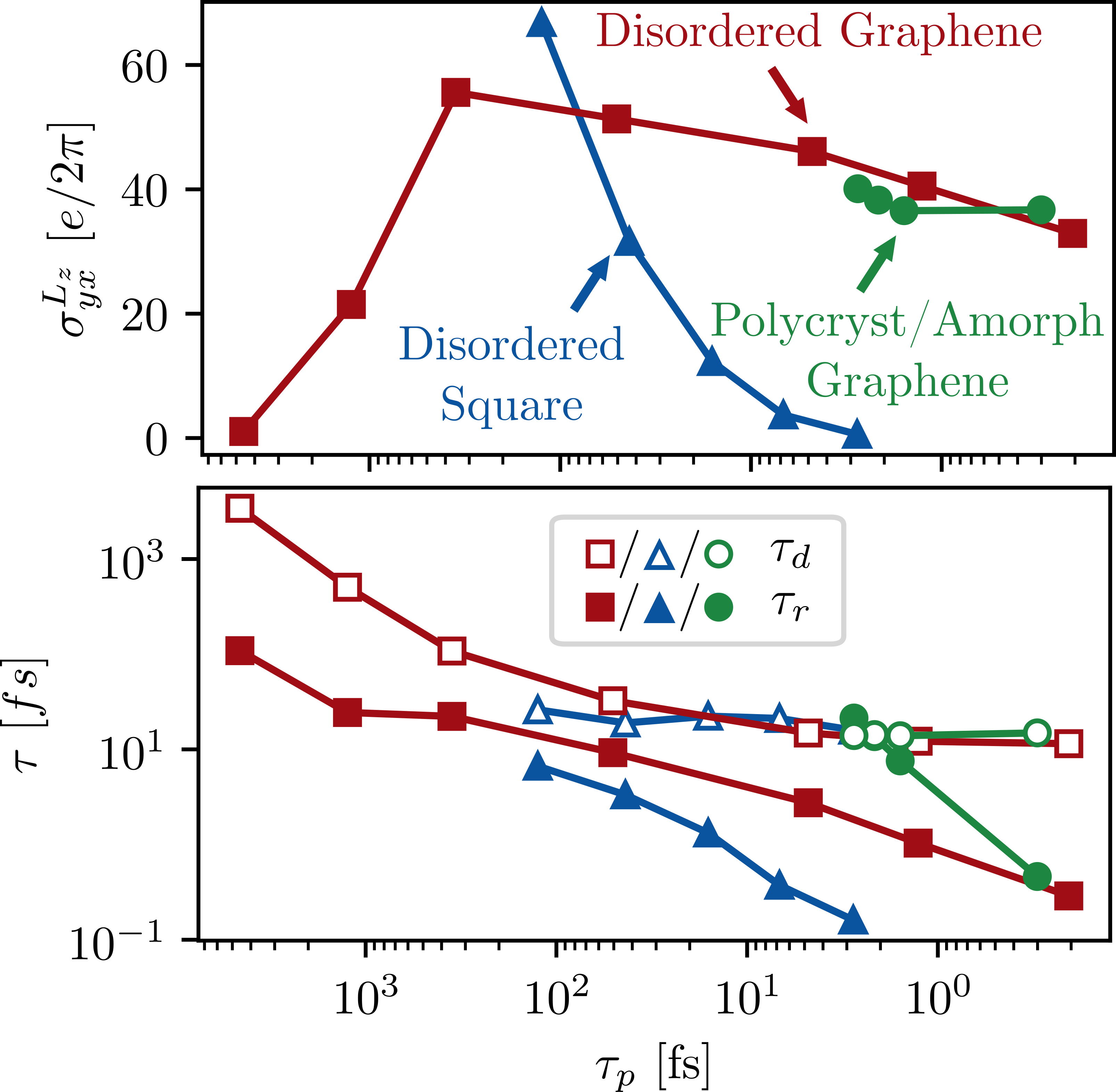}
\caption{\label{fig:summary_rise_decay_momentum} Top panel: saturated value of $\sigma_{yx}^{L_z}$ versus the momentum relaxation time for the square lattice (blue triangles), disordered graphene (red squares), and polycrystalline/amorphous graphene (green circles) at $\mu=0$. The values of the square lattice have been divided by two for visualization. Bottom panel: $\tau_r$ (closed symbols) and $\tau_d$ (open symbols) extracted from $C_{yx}^{L_z}$.}
\end{figure}

Next we examine the dependence of this phenomenon on disorder strength. As shown in the upper left panel of Fig. \ref{fig:square_graphene_amorph}, there is no Hall response in the clean limit ($W=0$), as expected. In the presence of disorder, a finite orbital response appears, and with increasing disorder strength the magnitude of $C_{yx}^{L_z}$ decreases monotonically as the ground state becomes more localized in real space, a factor that limits the formation of $\hat{v}_y$, the real-space spread of the ground state \cite{Romeral2025QuantumMetric}, and $\hat{L}_z$. This is also shown in the top panel of Fig. \ref{fig:summary_rise_decay_momentum}, where we plot the saturated value of $\sigma_{yx}^{L_z}$ as a function of momentum relaxation time, $\tau_p$, which is extracted from the longitudinal conductivity \cite{fan2021linear,suppmaterial,Guerrero2025}. Additionally, we plot $\tau_r$ and $\tau_d$ against $\tau_p$, in the bottom panel. We observe that $\tau_r$ decreases monotonically with increasing disorder strength, while $\tau_d$ remains approximately constant over the full range of disorder. Moreover, $\tau_r < \tau_d$ for all cases studied, and for the strongest disorder we find $\tau_r \ll \tau_d$. This is consistent with the picture that the phenomenon is driven by scattering -- stronger disorder leads to faster scattering and thus quicker formation of the orbital Hall response, yielding the scaling $\tau_r \propto \tau_p$. In contrast, the constant value of $\tau_d$ suggests that the loss of correlation is driven by a different mechanism. This will be discussed in more detail in the next section.

\textit{OHE in disordered graphene} - Next we consider the disorder-induced orbital Hall response in graphene with Anderson disorder. The time-dependent response is shown in the middle column of Fig.~\ref{fig:square_graphene_amorph}. As for the square lattice, there is no orbital response in clean graphene, while finite disorder strength gives rise to a finite $\sigma_{yx}^{L_z}$. However, graphene also exhibits a behavior that is distinct from the square lattice. In the top panel, the magnitude of the orbital-charge current correlator increases and eventually saturates with increasing disorder, in sharp contrast to the square lattice, whose magnitude of $C_{yx}^{L_z}$ decreases monotonically with $W$. 

This behavior also translates to the orbital Hall conductivity (bottom-middle panel of Fig. \ref{fig:square_graphene_amorph} and top panel of Fig. \ref{fig:summary_rise_decay_momentum}), whose magnitude exhibits nonmonotonic scaling with disorder strength. This may be attributed to the Dirac nature of graphene bands -- because large-angle scattering and backscattering are suppressed, more scattering events are required to build up the orbital Hall response. This leads to two regimes. In the weak-disorder regime, the scattering rate and associated dynamical Hall response is slow. Thus, increased disorder strength speeds up and enhances the magnitude of the Hall response. In the strong-disorder regime, localization sets in, suppressing the formation of OAM and reversing the trend of $\sigma_{yx}^{L_z}$ with disorder strength, similar to what is seen in the square lattice. Meanwhile, away from the graphene charge neutrality point, we recover the monotonic suppression of orbital Hall response with disorder strength similar to the square lattice case \cite{suppmaterial}.

The rise and decay times for graphene are shown as the red squares in the bottom panel of Fig.~\ref{fig:summary_rise_decay_momentum}. The rise time behaves similarly to that observed in the square lattice case with $\tau_r\propto\tau_p$, again highlighting that the formation of the orbital Hall response is driven by scattering events. In contrast, the decay time exhibits more complex behavior, with two distinct regimes: for weak disorder, $\tau_d \propto \tau_p$, while in the strong disorder limit $\tau_d$ saturates to a constant value, similar to the square lattice. This constant value coincides with a dephasing time, $\tau_{\varphi}$, that is inherent to our numerical approach and arises from the fact that the Chebyshev expansion is cut off at a finite number of polynomials, $M$. This yields $\tau_{\varphi} = \hbar M / (\pi \Delta E)$, where $\Delta E$ is the bandwidth of the disordered Hamiltonian \cite{fan2021linear,Weisse2006Kernel}. Here we have $\tau_{\varphi} \approx 21$ fs. The crossover in the behavior of $\tau_d$ occurs around the point $\tau_p = \tau_{\varphi}$, and we find empirically that it follows the relation $\tau_d \propto \tau_p + \tau_{\varphi}$. This suggests that the loss of correlation is driven by charge scattering in the diffusive  transport regime, $\tau_{\varphi} < \tau_p$ and is driven by dephasing in the coherent transport regime, $\tau_{\varphi}>\tau_p$.



It is important to highlight the magnitude of the orbital Hall conductivities in Fig.~\ref{fig:summary_rise_decay_momentum}. In general, our calculations find $\sigma_{yx}^{L_z}$ equals several tens in units of $e/2\pi$, which is at least one order of magnitude higher than other angular momentum Hall responses in Dirac materials, such as the OHE arising from the atomic basis in $\mathrm{MoS}_2$ or $\mathrm{WSe}_2$~\cite{canonico2020orbital}, or the typical values seen in the spin Hall effect~\cite{Sinova2015}. As shown in the SM~\cite{suppmaterial}, we see even larger values at higher chemical potentials.

\textit{OHE in polycrystalline and amorphous graphene} - Although Anderson disorder can reproduce some characteristic features of disordered graphene, it is not representative of real experimental systems. As such, we now consider the itinerant orbital Hall response in polycrystalline and amorphous graphene samples. These commonly appear in large-scale production methods, such as CVD~\cite{Zhang2013RevCVD, ChoiLarge2022}, where differently oriented graphene grains are separated by grain boundaries containing non-hexagonal rings. The extreme case corresponds to amorphous graphene, in which non-hexagonal rings are more uniformly distributed throughout the entire structure~\cite{Kotakoski2011,VanTuan2012,Toh2026Atomically,SK2026Quantum}. A small inset of these structures can be seen in the top row of Fig.~\ref{fig:square_graphene_amorph} and larger structures are shown in the SM \cite{suppmaterial}. We considered four geometries in total: three polycrystalline samples with average grain diameters of $24$, $20$, and $14$ nm, and one amorphous sample. Details on the sample generation approach and sample morphology can be found in the SM \cite{suppmaterial}.  All exhibit strong structural disorder, with localization lengths of $\xi_{\mathrm{loc}}\simeq4,5.7,7.4$ nm for polycrystalline samples and $1.1$ nm for the amorphous sample \cite{suppmaterial}.

The orbital Hall conductivity, in the bottom right panel of Fig.~\ref{fig:square_graphene_amorph}, is similar to the prior systems, increasing with time from zero and saturating at a finite value. Near charge neutrality ($\mu = 0$), shown by the solid lines, all polycrystalline and amorphous samples saturate at approximately the same value. This similarity arises because, at these energies, the states in all systems are localized around non-hexagonal rings -- at the grain boundaries in the polycrystalline samples and throughout the sample in the amorphous case~\cite{suppmaterial} -- leading to a universal orbital Hall response.

Away from charge neutrality at $\mu=0.1$~eV, as shown by the dashed lines, there is now a monotonic scaling with the degree of disorder -- the saturated value of $\sigma_{yx}^{L_z}$ decreases with decreasing grain size, and is smallest for the amorphous sample. At this chemical potential, states are distributed throughout the sample, both in the grains and at the non-hexagonal rings, and increasing localization by reducing grain size suppresses the orbital Hall response, similar to the square lattice and graphene at large disorder strength. Notably, the amorphous sample has approximately the same value of $\sigma_{yx}^{L_z}$ for both chemical potentials; this behavior originates from the presence of a broad energy range hosting strongly localized states \cite{VanTuan2012}.

We can also observe in Fig. \ref{fig:summary_rise_decay_momentum} the saturated values of the conductivity for $\mu=0$ eV. These values fall within the strong-disorder regime of graphene with Anderson disorder. As for the dynamics of the Hall response, the rise time scales with $\tau_p$ and the decay time saturates to the dephasing time, similar to the other systems. Finally, we note that the approximately constant long-time behavior of $\sigma_{yx}^{L_z}(t)$ is observed up to a simulation time of $2$~ps in the amorphous sample, which highlights the robustness of the orbital Hall response despite the strong charge localization \cite{suppmaterial}.

\textit{Summary and conclusions} - Using a time-dependent implementation of the Kubo formula, we demonstrate that disorder can generate an itinerant orbital Hall conductivity absent in clean systems, even in single-orbital models without atomic angular momentum or intrinsic orbital Hall conductivity. By accessing time-dependent charge–orbital current correlations, we confirm the role of scattering in generating such a response, whose dependence on the momentum relaxation time $\tau_p$ indicates an extrinsic origin. These findings echo prior works identifying skew-scattering and side-jump contributions to the OHE in gapped graphene and Weyl semimetals \cite{veneri2025extrinsic,Joy2026Skew,liu2024dominance}, but extend them as well as previous nonperturbative studies \cite{canonico2024orbital} to the case of gapless graphene and the trivial square lattice, and to strongly disordered and amorphous systems, suggesting the universality of this effect. Remarkably, we also observe a pronounced decoupling between charge and orbital transport, with polycrystalline and amorphous graphene sustaining long-lived itinerant orbital currents despite strong electronic localization. This is in contrast to the suppression of orbital currents predicted by calculations based on the atom-centered approximation of OAM \cite{tang2024role,rang2024orbital}. Our findings thus suggest that itinerant contributions may also play a substantial role in nonlocal orbital transport in multiorbital systems \cite{Barbosa2026Extrinsic}, consistent with recent graphene experiments \cite{SalvadorSanchez2024}. More broadly, our results identify structural disorder as a resource for generating and sustaining orbital Hall currents, as observed experimentally in disordered SrRuO3, amorphous Mo, and amorphous Mn \cite{peng2025unconventional,choi2026electron,wang2024}. They consequently establish amorphous and structurally complex materials as promising platforms for disorder-engineered orbitronics, long-range orbital transport, and orbital torque generation.
\begin{acknowledgments}
The authors JMR, LCA, AC and SR gratefully acknowledge the computational resources provided by the RES at BSC through the MareNostrum supercomputer under allocation FI-2025-3-0014. We acknowledge support from: Project PID2025-170043NB-I00, funded by MICIU/AEI/10.13039/501100011033 and by ERDF, EU; Project PID2025-170043NB-I00 funded by MICIU/AEI/10.13039/501100011033 and Ajuts per donar suport a l’activitat científica dels grups de recerca de Catalunya (2025 SGR 00240), Departament de Recerca i Universitats/Generalitat de Catalunya.

\end{acknowledgments}

\bibliography{bibliography.bib}

\begin{thebibliography}{68}%
\makeatletter
\providecommand \@ifxundefined [1]{%
 \@ifx{#1\undefined}
}%
\providecommand \@ifnum [1]{%
 \ifnum #1\expandafter \@firstoftwo
 \else \expandafter \@secondoftwo
 \fi
}%
\providecommand \@ifx [1]{%
 \ifx #1\expandafter \@firstoftwo
 \else \expandafter \@secondoftwo
 \fi
}%
\providecommand \natexlab [1]{#1}%
\providecommand \enquote  [1]{``#1''}%
\providecommand \bibnamefont  [1]{#1}%
\providecommand \bibfnamefont [1]{#1}%
\providecommand \citenamefont [1]{#1}%
\providecommand \href@noop [0]{\@secondoftwo}%
\providecommand \href [0]{\begingroup \@sanitize@url \@href}%
\providecommand \@href[1]{\@@startlink{#1}\@@href}%
\providecommand \@@href[1]{\endgroup#1\@@endlink}%
\providecommand \@sanitize@url [0]{\catcode `\\12\catcode `\$12\catcode `\&12\catcode `\#12\catcode `\^12\catcode `\_12\catcode `\%12\relax}%
\providecommand \@@startlink[1]{}%
\providecommand \@@endlink[0]{}%
\providecommand \url  [0]{\begingroup\@sanitize@url \@url }%
\providecommand \@url [1]{\endgroup\@href {#1}{\urlprefix }}%
\providecommand \urlprefix  [0]{URL }%
\providecommand \Eprint [0]{\href }%
\providecommand \doibase [0]{https://doi.org/}%
\providecommand \selectlanguage [0]{\@gobble}%
\providecommand \bibinfo  [0]{\@secondoftwo}%
\providecommand \bibfield  [0]{\@secondoftwo}%
\providecommand \translation [1]{[#1]}%
\providecommand \BibitemOpen [0]{}%
\providecommand \bibitemStop [0]{}%
\providecommand \bibitemNoStop [0]{.\EOS\space}%
\providecommand \EOS [0]{\spacefactor3000\relax}%
\providecommand \BibitemShut  [1]{\csname bibitem#1\endcsname}%
\let\auto@bib@innerbib\@empty
\bibitem [{\citenamefont {Ko}\ \emph {et~al.}(2026)\citenamefont {Ko}, \citenamefont {Jo}, \citenamefont {Kim}, \citenamefont {Oppeneer}, \citenamefont {Lee},\ and\ \citenamefont {Choi}}]{Ko2026Electron}%
  \BibitemOpen
  \bibfield  {author} {\bibinfo {author} {\bibfnamefont {K.-H.}\ \bibnamefont {Ko}}, \bibinfo {author} {\bibfnamefont {D.}~\bibnamefont {Jo}}, \bibinfo {author} {\bibfnamefont {S.}~\bibnamefont {Kim}}, \bibinfo {author} {\bibfnamefont {P.~M.}\ \bibnamefont {Oppeneer}}, \bibinfo {author} {\bibfnamefont {H.-W.}\ \bibnamefont {Lee}},\ and\ \bibinfo {author} {\bibfnamefont {G.-M.}\ \bibnamefont {Choi}},\ }\bibfield  {title} {\bibinfo {title} {Electron scattering effect on orbital transport in highly disordered nanocrystalline mn},\ }\href {https://doi.org/10.1021/acs.nanolett.6c03559} {\bibfield  {journal} {\bibinfo  {journal} {Nano Letters}\ }\textbf {\bibinfo {volume} {26}},\ \bibinfo {pages} {13501–13510} (\bibinfo {year} {2026})}\BibitemShut {NoStop}%
\bibitem [{\citenamefont {Go}\ \emph {et~al.}(2021)\citenamefont {Go}, \citenamefont {Jo}, \citenamefont {Lee}, \citenamefont {Kläui},\ and\ \citenamefont {Mokrousov}}]{go2021orbitronics}%
  \BibitemOpen
  \bibfield  {author} {\bibinfo {author} {\bibfnamefont {D.}~\bibnamefont {Go}}, \bibinfo {author} {\bibfnamefont {D.}~\bibnamefont {Jo}}, \bibinfo {author} {\bibfnamefont {H.-W.}\ \bibnamefont {Lee}}, \bibinfo {author} {\bibfnamefont {M.}~\bibnamefont {Kläui}},\ and\ \bibinfo {author} {\bibfnamefont {Y.}~\bibnamefont {Mokrousov}},\ }\bibfield  {title} {\bibinfo {title} {Orbitronics: Orbital currents in solids},\ }\href {https://doi.org/10.1209/0295-5075/ac2653} {\bibfield  {journal} {\bibinfo  {journal} {Europhysics Letters}\ }\textbf {\bibinfo {volume} {135}},\ \bibinfo {pages} {37001} (\bibinfo {year} {2021})}\BibitemShut {NoStop}%
\bibitem [{\citenamefont {Burgos~Atencia}\ \emph {et~al.}(2024)\citenamefont {Burgos~Atencia}, \citenamefont {Agarwal},\ and\ \citenamefont {Culcer}}]{burgos2024orbital}%
  \BibitemOpen
  \bibfield  {author} {\bibinfo {author} {\bibfnamefont {R.}~\bibnamefont {Burgos~Atencia}}, \bibinfo {author} {\bibfnamefont {A.}~\bibnamefont {Agarwal}},\ and\ \bibinfo {author} {\bibfnamefont {D.}~\bibnamefont {Culcer}},\ }\bibfield  {title} {\bibinfo {title} {Orbital angular momentum of bloch electrons: equilibrium formulation, magneto-electric phenomena, and the orbital hall effect},\ }\href {https://doi.org/10.1080/23746149.2024.2371972} {\bibfield  {journal} {\bibinfo  {journal} {Advances in Physics: X}\ }\textbf {\bibinfo {volume} {9}},\ \bibinfo {pages} {2371972} (\bibinfo {year} {2024})}\BibitemShut {NoStop}%
\bibitem [{\citenamefont {Ando}(2025)}]{ando2025orbitronics}%
  \BibitemOpen
  \bibfield  {author} {\bibinfo {author} {\bibfnamefont {K.}~\bibnamefont {Ando}},\ }\bibfield  {title} {\bibinfo {title} {Orbitronics: Harnessing orbital currents in solid-state devices},\ }\href {https://doi.org/10.7566/JPSJ.94.092001} {\bibfield  {journal} {\bibinfo  {journal} {Journal of the Physical Society of Japan}\ }\textbf {\bibinfo {volume} {94}},\ \bibinfo {pages} {092001} (\bibinfo {year} {2025})}\BibitemShut {NoStop}%
\bibitem [{\citenamefont {Cysne}\ \emph {et~al.}(2025)\citenamefont {Cysne}, \citenamefont {Canonico}, \citenamefont {Costa}, \citenamefont {Muniz},\ and\ \citenamefont {Rappoport}}]{cysne2025orbitronics}%
  \BibitemOpen
  \bibfield  {author} {\bibinfo {author} {\bibfnamefont {T.~P.}\ \bibnamefont {Cysne}}, \bibinfo {author} {\bibfnamefont {L.~M.}\ \bibnamefont {Canonico}}, \bibinfo {author} {\bibfnamefont {M.}~\bibnamefont {Costa}}, \bibinfo {author} {\bibfnamefont {R.~B.}\ \bibnamefont {Muniz}},\ and\ \bibinfo {author} {\bibfnamefont {T.~G.}\ \bibnamefont {Rappoport}},\ }\bibfield  {title} {\bibinfo {title} {Orbitronics in two-dimensional materials},\ }\href {https://doi.org/10.1038/s44306-025-00103-1} {\bibfield  {journal} {\bibinfo  {journal} {npj Spintronics}\ }\textbf {\bibinfo {volume} {3}},\ \bibinfo {pages} {39} (\bibinfo {year} {2025})}\BibitemShut {NoStop}%
\bibitem [{\citenamefont {Choi}\ \emph {et~al.}(2023)\citenamefont {Choi}, \citenamefont {Jo}, \citenamefont {Ko}, \citenamefont {Go}, \citenamefont {Kim}, \citenamefont {Park}, \citenamefont {Kim}, \citenamefont {Min}, \citenamefont {Choi},\ and\ \citenamefont {Lee}}]{choi2023observation}%
  \BibitemOpen
  \bibfield  {author} {\bibinfo {author} {\bibfnamefont {Y.-G.}\ \bibnamefont {Choi}}, \bibinfo {author} {\bibfnamefont {D.}~\bibnamefont {Jo}}, \bibinfo {author} {\bibfnamefont {K.-H.}\ \bibnamefont {Ko}}, \bibinfo {author} {\bibfnamefont {D.}~\bibnamefont {Go}}, \bibinfo {author} {\bibfnamefont {K.-H.}\ \bibnamefont {Kim}}, \bibinfo {author} {\bibfnamefont {H.~G.}\ \bibnamefont {Park}}, \bibinfo {author} {\bibfnamefont {C.}~\bibnamefont {Kim}}, \bibinfo {author} {\bibfnamefont {B.-C.}\ \bibnamefont {Min}}, \bibinfo {author} {\bibfnamefont {G.-M.}\ \bibnamefont {Choi}},\ and\ \bibinfo {author} {\bibfnamefont {H.-W.}\ \bibnamefont {Lee}},\ }\bibfield  {title} {\bibinfo {title} {Observation of the orbital hall effect in a light metal ti},\ }\href {https://doi.org/10.1038/s41586-023-06101-9} {\bibfield  {journal} {\bibinfo  {journal} {Nature}\ }\textbf {\bibinfo {volume} {619}},\ \bibinfo {pages} {52–56} (\bibinfo {year} {2023})}\BibitemShut {NoStop}%
\bibitem [{\citenamefont {Lyalin}\ \emph {et~al.}(2023)\citenamefont {Lyalin}, \citenamefont {Alikhah}, \citenamefont {Berritta}, \citenamefont {Oppeneer},\ and\ \citenamefont {Kawakami}}]{Lyalin2023}%
  \BibitemOpen
  \bibfield  {author} {\bibinfo {author} {\bibfnamefont {I.}~\bibnamefont {Lyalin}}, \bibinfo {author} {\bibfnamefont {S.}~\bibnamefont {Alikhah}}, \bibinfo {author} {\bibfnamefont {M.}~\bibnamefont {Berritta}}, \bibinfo {author} {\bibfnamefont {P.~M.}\ \bibnamefont {Oppeneer}},\ and\ \bibinfo {author} {\bibfnamefont {R.~K.}\ \bibnamefont {Kawakami}},\ }\bibfield  {title} {\bibinfo {title} {Magneto-optical detection of the orbital hall effect in chromium},\ }\href {https://doi.org/10.1103/PhysRevLett.131.156702} {\bibfield  {journal} {\bibinfo  {journal} {Phys. Rev. Lett.}\ }\textbf {\bibinfo {volume} {131}},\ \bibinfo {pages} {156702} (\bibinfo {year} {2023})}\BibitemShut {NoStop}%
\bibitem [{\citenamefont {Bernevig}\ \emph {et~al.}(2005)\citenamefont {Bernevig}, \citenamefont {Hughes},\ and\ \citenamefont {Zhang}}]{Bernevig2005}%
  \BibitemOpen
  \bibfield  {author} {\bibinfo {author} {\bibfnamefont {B.~A.}\ \bibnamefont {Bernevig}}, \bibinfo {author} {\bibfnamefont {T.~L.}\ \bibnamefont {Hughes}},\ and\ \bibinfo {author} {\bibfnamefont {S.-C.}\ \bibnamefont {Zhang}},\ }\bibfield  {title} {\bibinfo {title} {Orbitronics: The intrinsic orbital current in $p$-doped silicon},\ }\href {https://doi.org/10.1103/PhysRevLett.95.066601} {\bibfield  {journal} {\bibinfo  {journal} {Phys. Rev. Lett.}\ }\textbf {\bibinfo {volume} {95}},\ \bibinfo {pages} {066601} (\bibinfo {year} {2005})}\BibitemShut {NoStop}%
\bibitem [{\citenamefont {Go}\ \emph {et~al.}(2018)\citenamefont {Go}, \citenamefont {Jo}, \citenamefont {Kim},\ and\ \citenamefont {Lee}}]{Go2018}%
  \BibitemOpen
  \bibfield  {author} {\bibinfo {author} {\bibfnamefont {D.}~\bibnamefont {Go}}, \bibinfo {author} {\bibfnamefont {D.}~\bibnamefont {Jo}}, \bibinfo {author} {\bibfnamefont {C.}~\bibnamefont {Kim}},\ and\ \bibinfo {author} {\bibfnamefont {H.-W.}\ \bibnamefont {Lee}},\ }\bibfield  {title} {\bibinfo {title} {Intrinsic spin and orbital hall effects from orbital texture},\ }\href {https://doi.org/10.1103/PhysRevLett.121.086602} {\bibfield  {journal} {\bibinfo  {journal} {Phys. Rev. Lett.}\ }\textbf {\bibinfo {volume} {121}},\ \bibinfo {pages} {086602} (\bibinfo {year} {2018})}\BibitemShut {NoStop}%
\bibitem [{\citenamefont {Bhowal}\ and\ \citenamefont {Satpathy}(2020)}]{bhowal2020intrinsic}%
  \BibitemOpen
  \bibfield  {author} {\bibinfo {author} {\bibfnamefont {S.}~\bibnamefont {Bhowal}}\ and\ \bibinfo {author} {\bibfnamefont {S.}~\bibnamefont {Satpathy}},\ }\bibfield  {title} {\bibinfo {title} {Intrinsic orbital moment and prediction of a large orbital hall effect in two-dimensional transition metal dichalcogenides},\ }\href {https://doi.org/10.1103/PhysRevB.101.121112} {\bibfield  {journal} {\bibinfo  {journal} {Physical Review B}\ }\textbf {\bibinfo {volume} {101}},\ \bibinfo {pages} {121112} (\bibinfo {year} {2020})}\BibitemShut {NoStop}%
\bibitem [{\citenamefont {Canonico}\ \emph {et~al.}(2020)\citenamefont {Canonico}, \citenamefont {Cysne}, \citenamefont {Molina-Sanchez}, \citenamefont {Muniz},\ and\ \citenamefont {Rappoport}}]{canonico2020orbital}%
  \BibitemOpen
  \bibfield  {author} {\bibinfo {author} {\bibfnamefont {L.~M.}\ \bibnamefont {Canonico}}, \bibinfo {author} {\bibfnamefont {T.~P.}\ \bibnamefont {Cysne}}, \bibinfo {author} {\bibfnamefont {A.}~\bibnamefont {Molina-Sanchez}}, \bibinfo {author} {\bibfnamefont {R.~B.}\ \bibnamefont {Muniz}},\ and\ \bibinfo {author} {\bibfnamefont {T.~G.}\ \bibnamefont {Rappoport}},\ }\bibfield  {title} {\bibinfo {title} {Orbital hall insulating phase in transition metal dichalcogenide monolayers},\ }\href {https://doi.org/10.1103/PhysRevB.101.161409} {\bibfield  {journal} {\bibinfo  {journal} {Physical Review B}\ }\textbf {\bibinfo {volume} {101}},\ \bibinfo {pages} {161409} (\bibinfo {year} {2020})}\BibitemShut {NoStop}%
\bibitem [{\citenamefont {Go}\ and\ \citenamefont {Lee}(2020)}]{GoOrbitalTorque2020}%
  \BibitemOpen
  \bibfield  {author} {\bibinfo {author} {\bibfnamefont {D.}~\bibnamefont {Go}}\ and\ \bibinfo {author} {\bibfnamefont {H.-W.}\ \bibnamefont {Lee}},\ }\bibfield  {title} {\bibinfo {title} {Orbital torque: Torque generation by orbital current injection},\ }\href {https://doi.org/10.1103/PhysRevResearch.2.013177} {\bibfield  {journal} {\bibinfo  {journal} {Phys. Rev. Res.}\ }\textbf {\bibinfo {volume} {2}},\ \bibinfo {pages} {013177} (\bibinfo {year} {2020})}\BibitemShut {NoStop}%
\bibitem [{\citenamefont {Lee}\ \emph {et~al.}(2021)\citenamefont {Lee}, \citenamefont {Go}, \citenamefont {Park}, \citenamefont {Jeong}, \citenamefont {Ko}, \citenamefont {Yun}, \citenamefont {Jo}, \citenamefont {Lee}, \citenamefont {Go}, \citenamefont {Oh}, \citenamefont {Kim}, \citenamefont {Park}, \citenamefont {Min}, \citenamefont {Koo}, \citenamefont {Lee}, \citenamefont {Lee},\ and\ \citenamefont {Lee}}]{lee2021orbital}%
  \BibitemOpen
  \bibfield  {author} {\bibinfo {author} {\bibfnamefont {D.}~\bibnamefont {Lee}}, \bibinfo {author} {\bibfnamefont {D.}~\bibnamefont {Go}}, \bibinfo {author} {\bibfnamefont {H.-J.}\ \bibnamefont {Park}}, \bibinfo {author} {\bibfnamefont {W.}~\bibnamefont {Jeong}}, \bibinfo {author} {\bibfnamefont {H.-W.}\ \bibnamefont {Ko}}, \bibinfo {author} {\bibfnamefont {D.}~\bibnamefont {Yun}}, \bibinfo {author} {\bibfnamefont {D.}~\bibnamefont {Jo}}, \bibinfo {author} {\bibfnamefont {S.}~\bibnamefont {Lee}}, \bibinfo {author} {\bibfnamefont {G.}~\bibnamefont {Go}}, \bibinfo {author} {\bibfnamefont {J.~H.}\ \bibnamefont {Oh}}, \bibinfo {author} {\bibfnamefont {K.-J.}\ \bibnamefont {Kim}}, \bibinfo {author} {\bibfnamefont {B.-G.}\ \bibnamefont {Park}}, \bibinfo {author} {\bibfnamefont {B.-C.}\ \bibnamefont {Min}}, \bibinfo {author} {\bibfnamefont {H.~C.}\ \bibnamefont {Koo}}, \bibinfo {author} {\bibfnamefont {H.-W.}\ \bibnamefont {Lee}}, \bibinfo {author} {\bibfnamefont {O.}~\bibnamefont {Lee}},\ and\ \bibinfo {author} {\bibfnamefont {K.-J.}\ \bibnamefont {Lee}},\ }\bibfield  {title} {\bibinfo {title} {Orbital torque in magnetic bilayers},\ }\href {https://doi.org/10.1038/s41467-021-26650-9} {\bibfield  {journal} {\bibinfo  {journal} {Nature Communications}\ }\textbf {\bibinfo {volume} {12}},\ \bibinfo {pages} {6710} (\bibinfo {year} {2021})}\BibitemShut {NoStop}%
\bibitem [{\citenamefont {Hayashi}\ \emph {et~al.}(2023)\citenamefont {Hayashi}, \citenamefont {Jo}, \citenamefont {Go}, \citenamefont {Gao}, \citenamefont {Haku}, \citenamefont {Mokrousov}, \citenamefont {Lee},\ and\ \citenamefont {Ando}}]{hayashi2023observation}%
  \BibitemOpen
  \bibfield  {author} {\bibinfo {author} {\bibfnamefont {H.}~\bibnamefont {Hayashi}}, \bibinfo {author} {\bibfnamefont {D.}~\bibnamefont {Jo}}, \bibinfo {author} {\bibfnamefont {D.}~\bibnamefont {Go}}, \bibinfo {author} {\bibfnamefont {T.}~\bibnamefont {Gao}}, \bibinfo {author} {\bibfnamefont {S.}~\bibnamefont {Haku}}, \bibinfo {author} {\bibfnamefont {Y.}~\bibnamefont {Mokrousov}}, \bibinfo {author} {\bibfnamefont {H.-W.}\ \bibnamefont {Lee}},\ and\ \bibinfo {author} {\bibfnamefont {K.}~\bibnamefont {Ando}},\ }\bibfield  {title} {\bibinfo {title} {Observation of long-range orbital transport and giant orbital torque},\ }\href {https://doi.org/10.1038/s42005-023-01139-7} {\bibfield  {journal} {\bibinfo  {journal} {Communications Physics}\ }\textbf {\bibinfo {volume} {6}},\ \bibinfo {pages} {32} (\bibinfo {year} {2023})}\BibitemShut {NoStop}%
\bibitem [{\citenamefont {Gupta}\ \emph {et~al.}(2025)\citenamefont {Gupta}, \citenamefont {Bouard}, \citenamefont {Kammerbauer}, \citenamefont {Ledesma-Martin}, \citenamefont {Bose}, \citenamefont {Kononenko}, \citenamefont {Martin}, \citenamefont {Us{\'e}}, \citenamefont {Jakob}, \citenamefont {Drouard} \emph {et~al.}}]{gupta2025}%
  \BibitemOpen
  \bibfield  {author} {\bibinfo {author} {\bibfnamefont {R.}~\bibnamefont {Gupta}}, \bibinfo {author} {\bibfnamefont {C.}~\bibnamefont {Bouard}}, \bibinfo {author} {\bibfnamefont {F.}~\bibnamefont {Kammerbauer}}, \bibinfo {author} {\bibfnamefont {J.~O.}\ \bibnamefont {Ledesma-Martin}}, \bibinfo {author} {\bibfnamefont {A.}~\bibnamefont {Bose}}, \bibinfo {author} {\bibfnamefont {I.}~\bibnamefont {Kononenko}}, \bibinfo {author} {\bibfnamefont {S.}~\bibnamefont {Martin}}, \bibinfo {author} {\bibfnamefont {P.}~\bibnamefont {Us{\'e}}}, \bibinfo {author} {\bibfnamefont {G.}~\bibnamefont {Jakob}}, \bibinfo {author} {\bibfnamefont {M.}~\bibnamefont {Drouard}}, \emph {et~al.},\ }\bibfield  {title} {\bibinfo {title} {Harnessing orbital hall effect in spin-orbit torque mram},\ }\href@noop {} {\bibfield  {journal} {\bibinfo  {journal} {Nature Communications}\ }\textbf {\bibinfo {volume} {16}},\ \bibinfo {pages} {130} (\bibinfo {year} {2025})}\BibitemShut {NoStop}%
\bibitem [{\citenamefont {Santos}\ \emph {et~al.}(2023)\citenamefont {Santos}, \citenamefont {Abr\~ao}, \citenamefont {Go}, \citenamefont {de~Assis}, \citenamefont {Mokrousov}, \citenamefont {Mendes},\ and\ \citenamefont {Azevedo}}]{Santos2023}%
  \BibitemOpen
  \bibfield  {author} {\bibinfo {author} {\bibfnamefont {E.}~\bibnamefont {Santos}}, \bibinfo {author} {\bibfnamefont {J.}~\bibnamefont {Abr\~ao}}, \bibinfo {author} {\bibfnamefont {D.}~\bibnamefont {Go}}, \bibinfo {author} {\bibfnamefont {L.}~\bibnamefont {de~Assis}}, \bibinfo {author} {\bibfnamefont {Y.}~\bibnamefont {Mokrousov}}, \bibinfo {author} {\bibfnamefont {J.}~\bibnamefont {Mendes}},\ and\ \bibinfo {author} {\bibfnamefont {A.}~\bibnamefont {Azevedo}},\ }\bibfield  {title} {\bibinfo {title} {Inverse orbital torque via spin-orbital intertwined states},\ }\href {https://doi.org/10.1103/PhysRevApplied.19.014069} {\bibfield  {journal} {\bibinfo  {journal} {Phys. Rev. Appl.}\ }\textbf {\bibinfo {volume} {19}},\ \bibinfo {pages} {014069} (\bibinfo {year} {2023})}\BibitemShut {NoStop}%
\bibitem [{\citenamefont {Santos}\ \emph {et~al.}(2024)\citenamefont {Santos}, \citenamefont {Abrão}, \citenamefont {Vieira}, \citenamefont {Mendes}, \citenamefont {Rodríguez-Suárez},\ and\ \citenamefont {Azevedo}}]{santos2024exploring}%
  \BibitemOpen
  \bibfield  {author} {\bibinfo {author} {\bibfnamefont {E.}~\bibnamefont {Santos}}, \bibinfo {author} {\bibfnamefont {J.~E.}\ \bibnamefont {Abrão}}, \bibinfo {author} {\bibfnamefont {A.~S.}\ \bibnamefont {Vieira}}, \bibinfo {author} {\bibfnamefont {J.~B.~S.}\ \bibnamefont {Mendes}}, \bibinfo {author} {\bibfnamefont {R.~L.}\ \bibnamefont {Rodríguez-Suárez}},\ and\ \bibinfo {author} {\bibfnamefont {A.}~\bibnamefont {Azevedo}},\ }\bibfield  {title} {\bibinfo {title} {Exploring orbital-charge conversion mediated by interfaces with $\mathrm{Cu}{\mathrm{o}}_{x}$ through spin-orbital pumping},\ }\href {https://doi.org/10.1103/PhysRevB.109.014420} {\bibfield  {journal} {\bibinfo  {journal} {Physical Review B}\ }\textbf {\bibinfo {volume} {109}},\ \bibinfo {pages} {014420} (\bibinfo {year} {2024})}\BibitemShut {NoStop}%
\bibitem [{\citenamefont {Hayashi}\ \emph {et~al.}(2024)\citenamefont {Hayashi}, \citenamefont {Go}, \citenamefont {Haku}, \citenamefont {Mokrousov},\ and\ \citenamefont {Ando}}]{hayashi2024observation}%
  \BibitemOpen
  \bibfield  {author} {\bibinfo {author} {\bibfnamefont {H.}~\bibnamefont {Hayashi}}, \bibinfo {author} {\bibfnamefont {D.}~\bibnamefont {Go}}, \bibinfo {author} {\bibfnamefont {S.}~\bibnamefont {Haku}}, \bibinfo {author} {\bibfnamefont {Y.}~\bibnamefont {Mokrousov}},\ and\ \bibinfo {author} {\bibfnamefont {K.}~\bibnamefont {Ando}},\ }\bibfield  {title} {\bibinfo {title} {Observation of orbital pumping},\ }\href {https://doi.org/10.1038/s41928-024-01193-1} {\bibfield  {journal} {\bibinfo  {journal} {Nature Electronics}\ }\textbf {\bibinfo {volume} {7}},\ \bibinfo {pages} {646–652} (\bibinfo {year} {2024})}\BibitemShut {NoStop}%
\bibitem [{\citenamefont {Abrão}\ \emph {et~al.}(2025)\citenamefont {Abrão}, \citenamefont {Santos}, \citenamefont {Costa}, \citenamefont {Santos}, \citenamefont {Mendes},\ and\ \citenamefont {Azevedo}}]{abrao2025anomalous}%
  \BibitemOpen
  \bibfield  {author} {\bibinfo {author} {\bibfnamefont {J.}~\bibnamefont {Abrão}}, \bibinfo {author} {\bibfnamefont {E.}~\bibnamefont {Santos}}, \bibinfo {author} {\bibfnamefont {J.}~\bibnamefont {Costa}}, \bibinfo {author} {\bibfnamefont {J.}~\bibnamefont {Santos}}, \bibinfo {author} {\bibfnamefont {J.}~\bibnamefont {Mendes}},\ and\ \bibinfo {author} {\bibfnamefont {A.}~\bibnamefont {Azevedo}},\ }\bibfield  {title} {\bibinfo {title} {Anomalous spin and orbital hall phenomena in antiferromagnetic systems},\ }\href {https://doi.org/10.1103/PhysRevLett.134.026702} {\bibfield  {journal} {\bibinfo  {journal} {Physical Review Letters}\ }\textbf {\bibinfo {volume} {134}},\ \bibinfo {pages} {026702} (\bibinfo {year} {2025})}\BibitemShut {NoStop}%
\bibitem [{\citenamefont {Costa}\ \emph {et~al.}(2025)\citenamefont {Costa}, \citenamefont {Santos}, \citenamefont {Mendes},\ and\ \citenamefont {Azevedo}}]{costa2025dominance}%
  \BibitemOpen
  \bibfield  {author} {\bibinfo {author} {\bibfnamefont {J.~L.}\ \bibnamefont {Costa}}, \bibinfo {author} {\bibfnamefont {E.}~\bibnamefont {Santos}}, \bibinfo {author} {\bibfnamefont {J.~B.~S.}\ \bibnamefont {Mendes}},\ and\ \bibinfo {author} {\bibfnamefont {A.}~\bibnamefont {Azevedo}},\ }\bibfield  {title} {\bibinfo {title} {Dominance of the orbital hall effect over spin in transition metal heterostructures},\ }\href {https://doi.org/10.1103/tbdz-jc6k} {\bibfield  {journal} {\bibinfo  {journal} {Physical Review B}\ }\textbf {\bibinfo {volume} {112}},\ \bibinfo {pages} {054443} (\bibinfo {year} {2025})}\BibitemShut {NoStop}%
\bibitem [{\citenamefont {Seifert}\ \emph {et~al.}(2023)\citenamefont {Seifert}, \citenamefont {Go}, \citenamefont {Hayashi}, \citenamefont {Rouzegar}, \citenamefont {Freimuth}, \citenamefont {Ando}, \citenamefont {Mokrousov},\ and\ \citenamefont {Kampfrath}}]{seifert2023}%
  \BibitemOpen
  \bibfield  {author} {\bibinfo {author} {\bibfnamefont {T.~S.}\ \bibnamefont {Seifert}}, \bibinfo {author} {\bibfnamefont {D.}~\bibnamefont {Go}}, \bibinfo {author} {\bibfnamefont {H.}~\bibnamefont {Hayashi}}, \bibinfo {author} {\bibfnamefont {R.}~\bibnamefont {Rouzegar}}, \bibinfo {author} {\bibfnamefont {F.}~\bibnamefont {Freimuth}}, \bibinfo {author} {\bibfnamefont {K.}~\bibnamefont {Ando}}, \bibinfo {author} {\bibfnamefont {Y.}~\bibnamefont {Mokrousov}},\ and\ \bibinfo {author} {\bibfnamefont {T.}~\bibnamefont {Kampfrath}},\ }\bibfield  {title} {\bibinfo {title} {Time-domain observation of ballistic orbital-angular-momentum currents with giant relaxation length in tungsten},\ }\href {https://doi.org/10.1038/s41565-023-01470-8} {\bibfield  {journal} {\bibinfo  {journal} {Nature Nanotechnology}\ }\textbf {\bibinfo {volume} {18}},\ \bibinfo {pages} {1132–1138} (\bibinfo {year} {2023})}\BibitemShut {NoStop}%
\bibitem [{\citenamefont {Xu}\ \emph {et~al.}(2024)\citenamefont {Xu}, \citenamefont {Zhang}, \citenamefont {Fert}, \citenamefont {Jaffres}, \citenamefont {Liu}, \citenamefont {Xu}, \citenamefont {Jiang}, \citenamefont {Cheng},\ and\ \citenamefont {Zhao}}]{xu2024orbitronics}%
  \BibitemOpen
  \bibfield  {author} {\bibinfo {author} {\bibfnamefont {Y.}~\bibnamefont {Xu}}, \bibinfo {author} {\bibfnamefont {F.}~\bibnamefont {Zhang}}, \bibinfo {author} {\bibfnamefont {A.}~\bibnamefont {Fert}}, \bibinfo {author} {\bibfnamefont {H.-Y.}\ \bibnamefont {Jaffres}}, \bibinfo {author} {\bibfnamefont {Y.}~\bibnamefont {Liu}}, \bibinfo {author} {\bibfnamefont {R.}~\bibnamefont {Xu}}, \bibinfo {author} {\bibfnamefont {Y.}~\bibnamefont {Jiang}}, \bibinfo {author} {\bibfnamefont {H.}~\bibnamefont {Cheng}},\ and\ \bibinfo {author} {\bibfnamefont {W.}~\bibnamefont {Zhao}},\ }\bibfield  {title} {\bibinfo {title} {Orbitronics: light-induced orbital currents in ni studied by terahertz emission experiments},\ }\href {https://doi.org/10.1038/s41467-024-46405-6} {\bibfield  {journal} {\bibinfo  {journal} {Nature Communications}\ }\textbf {\bibinfo {volume} {15}},\ \bibinfo {pages} {2043} (\bibinfo {year} {2024})}\BibitemShut {NoStop}%
\bibitem [{\citenamefont {Liu}\ \emph {et~al.}(2024)\citenamefont {Liu}, \citenamefont {Xu}, \citenamefont {Fert}, \citenamefont {Jaffrès}, \citenamefont {Nie}, \citenamefont {Eimer}, \citenamefont {Zhang},\ and\ \citenamefont {Zhao}}]{liu2024efficient}%
  \BibitemOpen
  \bibfield  {author} {\bibinfo {author} {\bibfnamefont {Y.}~\bibnamefont {Liu}}, \bibinfo {author} {\bibfnamefont {Y.}~\bibnamefont {Xu}}, \bibinfo {author} {\bibfnamefont {A.}~\bibnamefont {Fert}}, \bibinfo {author} {\bibfnamefont {H.-Y.}\ \bibnamefont {Jaffrès}}, \bibinfo {author} {\bibfnamefont {T.}~\bibnamefont {Nie}}, \bibinfo {author} {\bibfnamefont {S.}~\bibnamefont {Eimer}}, \bibinfo {author} {\bibfnamefont {X.}~\bibnamefont {Zhang}},\ and\ \bibinfo {author} {\bibfnamefont {W.}~\bibnamefont {Zhao}},\ }\bibfield  {title} {\bibinfo {title} {Efficient orbitronic terahertz emission based on copt alloy},\ }\href {https://doi.org/10.1002/adma.202404174} {\bibfield  {journal} {\bibinfo  {journal} {Advanced Materials}\ }\textbf {\bibinfo {volume} {36}},\ \bibinfo {pages} {2404174} (\bibinfo {year} {2024})}\BibitemShut {NoStop}%
\bibitem [{\citenamefont {Gao}\ \emph {et~al.}(2025)\citenamefont {Gao}, \citenamefont {Liao}, \citenamefont {Isshiki}, \citenamefont {Budai}, \citenamefont {Kim}, \citenamefont {Lee}, \citenamefont {Lee}, \citenamefont {Go}, \citenamefont {Mokrousov}, \citenamefont {Miwa},\ and\ \citenamefont {Otani}}]{gao2025nonlocal}%
  \BibitemOpen
  \bibfield  {author} {\bibinfo {author} {\bibfnamefont {W.}~\bibnamefont {Gao}}, \bibinfo {author} {\bibfnamefont {L.}~\bibnamefont {Liao}}, \bibinfo {author} {\bibfnamefont {H.}~\bibnamefont {Isshiki}}, \bibinfo {author} {\bibfnamefont {N.}~\bibnamefont {Budai}}, \bibinfo {author} {\bibfnamefont {J.}~\bibnamefont {Kim}}, \bibinfo {author} {\bibfnamefont {H.-W.}\ \bibnamefont {Lee}}, \bibinfo {author} {\bibfnamefont {K.-J.}\ \bibnamefont {Lee}}, \bibinfo {author} {\bibfnamefont {D.}~\bibnamefont {Go}}, \bibinfo {author} {\bibfnamefont {Y.}~\bibnamefont {Mokrousov}}, \bibinfo {author} {\bibfnamefont {S.}~\bibnamefont {Miwa}},\ and\ \bibinfo {author} {\bibfnamefont {Y.}~\bibnamefont {Otani}},\ }\bibfield  {title} {\bibinfo {title} {Nonlocal electrical detection of reciprocal orbital edelstein effect},\ }\href {https://doi.org/10.1038/s41467-025-61602-7} {\bibfield  {journal} {\bibinfo  {journal} {Nature Communications}\ }\textbf {\bibinfo {volume} {16}},\ \bibinfo {pages} {6380} (\bibinfo {year} {2025})}\BibitemShut {NoStop}%
\bibitem [{\citenamefont {Wang}\ \emph {et~al.}(2024)\citenamefont {Wang}, \citenamefont {Meng}, \citenamefont {Yao}, \citenamefont {Lin}, \citenamefont {Bai}, \citenamefont {Ning}, \citenamefont {Li}, \citenamefont {Zhang}, \citenamefont {Nie}, \citenamefont {Shi} \emph {et~al.}}]{wang2024}%
  \BibitemOpen
  \bibfield  {author} {\bibinfo {author} {\bibfnamefont {X.}~\bibnamefont {Wang}}, \bibinfo {author} {\bibfnamefont {A.}~\bibnamefont {Meng}}, \bibinfo {author} {\bibfnamefont {Y.}~\bibnamefont {Yao}}, \bibinfo {author} {\bibfnamefont {F.}~\bibnamefont {Lin}}, \bibinfo {author} {\bibfnamefont {Y.}~\bibnamefont {Bai}}, \bibinfo {author} {\bibfnamefont {X.}~\bibnamefont {Ning}}, \bibinfo {author} {\bibfnamefont {B.}~\bibnamefont {Li}}, \bibinfo {author} {\bibfnamefont {Y.}~\bibnamefont {Zhang}}, \bibinfo {author} {\bibfnamefont {T.}~\bibnamefont {Nie}}, \bibinfo {author} {\bibfnamefont {S.}~\bibnamefont {Shi}}, \emph {et~al.},\ }\bibfield  {title} {\bibinfo {title} {Large field-like spin--orbit torque and enhanced magnetization switching efficiency utilizing amorphous mo},\ }\href@noop {} {\bibfield  {journal} {\bibinfo  {journal} {Nano Letters}\ }\textbf {\bibinfo {volume} {24}},\ \bibinfo {pages} {6931} (\bibinfo {year} {2024})}\BibitemShut {NoStop}%
\bibitem [{\citenamefont {Peng}\ \emph {et~al.}(2025)\citenamefont {Peng}, \citenamefont {Zheng}, \citenamefont {Li}, \citenamefont {Lao}, \citenamefont {Han}, \citenamefont {Liao}, \citenamefont {Zheng}, \citenamefont {Yang}, \citenamefont {Yu}, \citenamefont {Liu}, \citenamefont {Sun}, \citenamefont {Chen}, \citenamefont {Peng}, \citenamefont {Zhao}, \citenamefont {Li},\ and\ \citenamefont {Wang}}]{peng2025unconventional}%
  \BibitemOpen
  \bibfield  {author} {\bibinfo {author} {\bibfnamefont {S.}~\bibnamefont {Peng}}, \bibinfo {author} {\bibfnamefont {X.}~\bibnamefont {Zheng}}, \bibinfo {author} {\bibfnamefont {S.}~\bibnamefont {Li}}, \bibinfo {author} {\bibfnamefont {B.}~\bibnamefont {Lao}}, \bibinfo {author} {\bibfnamefont {Y.}~\bibnamefont {Han}}, \bibinfo {author} {\bibfnamefont {Z.}~\bibnamefont {Liao}}, \bibinfo {author} {\bibfnamefont {H.}~\bibnamefont {Zheng}}, \bibinfo {author} {\bibfnamefont {Y.}~\bibnamefont {Yang}}, \bibinfo {author} {\bibfnamefont {T.}~\bibnamefont {Yu}}, \bibinfo {author} {\bibfnamefont {P.}~\bibnamefont {Liu}}, \bibinfo {author} {\bibfnamefont {Y.}~\bibnamefont {Sun}}, \bibinfo {author} {\bibfnamefont {X.-Q.}\ \bibnamefont {Chen}}, \bibinfo {author} {\bibfnamefont {S.}~\bibnamefont {Peng}}, \bibinfo {author} {\bibfnamefont {W.}~\bibnamefont {Zhao}}, \bibinfo {author} {\bibfnamefont {R.-W.}\ \bibnamefont {Li}},\ and\ \bibinfo {author} {\bibfnamefont {Z.}~\bibnamefont {Wang}},\ }\bibfield  {title} {\bibinfo {title} {Unconventional scaling of the orbital hall effect},\ }\href {https://doi.org/10.1038/s41563-025-02326-3} {\bibfield  {journal} {\bibinfo  {journal} {Nature Materials}\ }\textbf {\bibinfo {volume} {24}},\ \bibinfo {pages} {1749–1755} (\bibinfo {year} {2025})}\BibitemShut {NoStop}%
\bibitem [{\citenamefont {Choi}\ \emph {et~al.}(2026)\citenamefont {Choi}, \citenamefont {Ko}, \citenamefont {Jo} \emph {et~al.}}]{choi2026electron}%
  \BibitemOpen
  \bibfield  {author} {\bibinfo {author} {\bibfnamefont {G.-M.}\ \bibnamefont {Choi}}, \bibinfo {author} {\bibfnamefont {K.-H.}\ \bibnamefont {Ko}}, \bibinfo {author} {\bibfnamefont {D.}~\bibnamefont {Jo}}, \emph {et~al.},\ }\bibfield  {title} {\bibinfo {title} {Electron scattering enhances orbital transport in amorphous mn},\ }\bibfield  {journal} {\bibinfo  {journal} {Research Square}\ }\href {https://doi.org/10.21203/rs.3.rs-9248804/v1} {10.21203/rs.3.rs-9248804/v1} (\bibinfo {year} {2026}),\ \bibinfo {note} {preprint (Version 1)}\BibitemShut {NoStop}%
\bibitem [{\citenamefont {Tang}\ and\ \citenamefont {Bauer}(2024)}]{tang2024role}%
  \BibitemOpen
  \bibfield  {author} {\bibinfo {author} {\bibfnamefont {P.}~\bibnamefont {Tang}}\ and\ \bibinfo {author} {\bibfnamefont {G.~E.}\ \bibnamefont {Bauer}},\ }\bibfield  {title} {\bibinfo {title} {Role of disorder in the intrinsic orbital hall effect},\ }\href {https://doi.org/10.1103/PhysRevLett.133.186302} {\bibfield  {journal} {\bibinfo  {journal} {Physical Review Letters}\ }\textbf {\bibinfo {volume} {133}},\ \bibinfo {pages} {186302} (\bibinfo {year} {2024})}\BibitemShut {NoStop}%
\bibitem [{\citenamefont {Rang}\ and\ \citenamefont {Kelly}(2024)}]{rang2024orbital}%
  \BibitemOpen
  \bibfield  {author} {\bibinfo {author} {\bibfnamefont {M.}~\bibnamefont {Rang}}\ and\ \bibinfo {author} {\bibfnamefont {P.~J.}\ \bibnamefont {Kelly}},\ }\bibfield  {title} {\bibinfo {title} {Orbital relaxation length from first-principles scattering calculations},\ }\href {https://doi.org/10.1103/PhysRevB.109.214427} {\bibfield  {journal} {\bibinfo  {journal} {Physical Review B}\ }\textbf {\bibinfo {volume} {109}},\ \bibinfo {pages} {214427} (\bibinfo {year} {2024})}\BibitemShut {NoStop}%
\bibitem [{\citenamefont {Walz}\ \emph {et~al.}(2014)\citenamefont {Walz}, \citenamefont {Wilhelm},\ and\ \citenamefont {Evers}}]{Walz2014}%
  \BibitemOpen
  \bibfield  {author} {\bibinfo {author} {\bibfnamefont {M.}~\bibnamefont {Walz}}, \bibinfo {author} {\bibfnamefont {J.}~\bibnamefont {Wilhelm}},\ and\ \bibinfo {author} {\bibfnamefont {F.}~\bibnamefont {Evers}},\ }\bibfield  {title} {\bibinfo {title} {Current patterns and orbital magnetism in mesoscopic dc transport},\ }\href {https://doi.org/10.1103/PhysRevLett.113.136602} {\bibfield  {journal} {\bibinfo  {journal} {Phys. Rev. Lett.}\ }\textbf {\bibinfo {volume} {113}},\ \bibinfo {pages} {136602} (\bibinfo {year} {2014})}\BibitemShut {NoStop}%
\bibitem [{\citenamefont {Veneri}\ \emph {et~al.}(2025)\citenamefont {Veneri}, \citenamefont {Rappoport},\ and\ \citenamefont {Ferreira}}]{veneri2025extrinsic}%
  \BibitemOpen
  \bibfield  {author} {\bibinfo {author} {\bibfnamefont {A.}~\bibnamefont {Veneri}}, \bibinfo {author} {\bibfnamefont {T.~G.}\ \bibnamefont {Rappoport}},\ and\ \bibinfo {author} {\bibfnamefont {A.}~\bibnamefont {Ferreira}},\ }\bibfield  {title} {\bibinfo {title} {Extrinsic orbital hall effect: Orbital skew scattering and crossover between diffusive and intrinsic orbital transport},\ }\href {https://doi.org/10.1103/PhysRevLett.134.136201} {\bibfield  {journal} {\bibinfo  {journal} {Physical Review Letters}\ }\textbf {\bibinfo {volume} {134}},\ \bibinfo {pages} {136201} (\bibinfo {year} {2025})}\BibitemShut {NoStop}%
\bibitem [{\citenamefont {Joy}\ \emph {et~al.}(2026)\citenamefont {Joy}, \citenamefont {Pandey}, \citenamefont {Atencia},\ and\ \citenamefont {Bhalla}}]{Joy2026Skew}%
  \BibitemOpen
  \bibfield  {author} {\bibinfo {author} {\bibfnamefont {D.}~\bibnamefont {Joy}}, \bibinfo {author} {\bibfnamefont {V.}~\bibnamefont {Pandey}}, \bibinfo {author} {\bibfnamefont {R.~B.}\ \bibnamefont {Atencia}},\ and\ \bibinfo {author} {\bibfnamefont {P.}~\bibnamefont {Bhalla}},\ }\bibfield  {title} {\bibinfo {title} {Skew scattering induced contribution to orbital hall response},\ }\bibfield  {journal} {\bibinfo  {journal} {arXiv}\ }\href {https://doi.org/10.48550/arXiv.2607.24671} {10.48550/arXiv.2607.24671} (\bibinfo {year} {2026}),\ \bibinfo {note} {arXiv:2607.24671-cond-mat.mes-hall}\BibitemShut {NoStop}%
\bibitem [{\citenamefont {Vidarte}\ \emph {et~al.}(2026)\citenamefont {Vidarte}, \citenamefont {Veiga}, \citenamefont {Lopes}, \citenamefont {Cardias}, \citenamefont {Ferreira}, \citenamefont {Cysne},\ and\ \citenamefont {Rappoport}}]{Vidarte2026real}%
  \BibitemOpen
  \bibfield  {author} {\bibinfo {author} {\bibfnamefont {K.~J.~U.}\ \bibnamefont {Vidarte}}, \bibinfo {author} {\bibfnamefont {H.~P.}\ \bibnamefont {Veiga}}, \bibinfo {author} {\bibfnamefont {J.~M. V.~P.}\ \bibnamefont {Lopes}}, \bibinfo {author} {\bibfnamefont {R.}~\bibnamefont {Cardias}}, \bibinfo {author} {\bibfnamefont {A.}~\bibnamefont {Ferreira}}, \bibinfo {author} {\bibfnamefont {T.~P.}\ \bibnamefont {Cysne}},\ and\ \bibinfo {author} {\bibfnamefont {T.~G.}\ \bibnamefont {Rappoport}},\ }\bibfield  {title} {\bibinfo {title} {Real-space spectral approach to orbital magnetization},\ }\href {https://doi.org/10.1103/bhg8-c3rw} {\bibfield  {journal} {\bibinfo  {journal} {Physical Review B}\ }\textbf {\bibinfo {volume} {113}},\ \bibinfo {pages} {224438} (\bibinfo {year} {2026})}\BibitemShut {NoStop}%
\bibitem [{\citenamefont {Canonico}\ \emph {et~al.}(2024)\citenamefont {Canonico}, \citenamefont {Garcia},\ and\ \citenamefont {Roche}}]{canonico2024orbital}%
  \BibitemOpen
  \bibfield  {author} {\bibinfo {author} {\bibfnamefont {L.~M.}\ \bibnamefont {Canonico}}, \bibinfo {author} {\bibfnamefont {J.~H.}\ \bibnamefont {Garcia}},\ and\ \bibinfo {author} {\bibfnamefont {S.}~\bibnamefont {Roche}},\ }\bibfield  {title} {\bibinfo {title} {Orbital hall responses in disordered topological materials},\ }\href {https://doi.org/10.1103/PhysRevB.110.L140201} {\bibfield  {journal} {\bibinfo  {journal} {Physical Review B}\ }\textbf {\bibinfo {volume} {110}},\ \bibinfo {pages} {L140201} (\bibinfo {year} {2024})}\BibitemShut {NoStop}%
\bibitem [{\citenamefont {Barbosa}\ \emph {et~al.}(2026)\citenamefont {Barbosa}, \citenamefont {Lee},\ and\ \citenamefont {Rappoport}}]{Barbosa2026Extrinsic}%
  \BibitemOpen
  \bibfield  {author} {\bibinfo {author} {\bibfnamefont {A.~L.~R.}\ \bibnamefont {Barbosa}}, \bibinfo {author} {\bibfnamefont {H.-W.}\ \bibnamefont {Lee}},\ and\ \bibinfo {author} {\bibfnamefont {T.~G.}\ \bibnamefont {Rappoport}},\ }\bibfield  {title} {\bibinfo {title} {Extrinsic orbital hall effect and orbital relaxation in mesoscopic devices},\ }\href {https://doi.org/10.1103/79yg-jg9p} {\bibfield  {journal} {\bibinfo  {journal} {Physical Review B}\ }\textbf {\bibinfo {volume} {114}},\ \bibinfo {pages} {214401} (\bibinfo {year} {2026})}\BibitemShut {NoStop}%
\bibitem [{\citenamefont {Aversa}\ and\ \citenamefont {Sipe}(1995)}]{Aversa1995}%
  \BibitemOpen
  \bibfield  {author} {\bibinfo {author} {\bibfnamefont {C.}~\bibnamefont {Aversa}}\ and\ \bibinfo {author} {\bibfnamefont {J.~E.}\ \bibnamefont {Sipe}},\ }\bibfield  {title} {\bibinfo {title} {Nonlinear optical susceptibilities of semiconductors: Results with a length-gauge analysis},\ }\href {https://doi.org/10.1103/physrevb.52.14636} {\bibfield  {journal} {\bibinfo  {journal} {Physical Review B}\ }\textbf {\bibinfo {volume} {52}},\ \bibinfo {pages} {14636–14645} (\bibinfo {year} {1995})}\BibitemShut {NoStop}%
\bibitem [{\citenamefont {Zubarev}(1960)}]{zubarev1960}%
  \BibitemOpen
  \bibfield  {author} {\bibinfo {author} {\bibfnamefont {D.~N.}\ \bibnamefont {Zubarev}},\ }\bibfield  {title} {\bibinfo {title} {Double-time green functions in statistical physics},\ }\href@noop {} {\bibfield  {journal} {\bibinfo  {journal} {Soviet Physics Uspekhi}\ }\textbf {\bibinfo {volume} {3}},\ \bibinfo {pages} {320} (\bibinfo {year} {1960})}\BibitemShut {NoStop}%
\bibitem [{\citenamefont {Romeral}\ \emph {et~al.}(2025)\citenamefont {Romeral}, \citenamefont {Cummings},\ and\ \citenamefont {Roche}}]{Romeral2025QuantumMetric}%
  \BibitemOpen
  \bibfield  {author} {\bibinfo {author} {\bibfnamefont {J.~M.}\ \bibnamefont {Romeral}}, \bibinfo {author} {\bibfnamefont {A.~W.}\ \bibnamefont {Cummings}},\ and\ \bibinfo {author} {\bibfnamefont {S.}~\bibnamefont {Roche}},\ }\bibfield  {title} {\bibinfo {title} {Scaling of the integrated quantum metric in disordered topological phases},\ }\href {https://doi.org/10.1103/physrevb.111.134201} {\bibfield  {journal} {\bibinfo  {journal} {Physical Review B}\ }\textbf {\bibinfo {volume} {111}},\ \bibinfo {pages} {134201} (\bibinfo {year} {2025})}\BibitemShut {NoStop}%
\bibitem [{\citenamefont {Veiga}\ \emph {et~al.}(2024)\citenamefont {Veiga}, \citenamefont {João}, \citenamefont {Alendouro~Pinho}, \citenamefont {Santos~Pires},\ and\ \citenamefont {Lopes}}]{veiga2024unambiguous}%
  \BibitemOpen
  \bibfield  {author} {\bibinfo {author} {\bibfnamefont {H.}~\bibnamefont {Veiga}}, \bibinfo {author} {\bibfnamefont {S.~M.}\ \bibnamefont {João}}, \bibinfo {author} {\bibfnamefont {J.~M.}\ \bibnamefont {Alendouro~Pinho}}, \bibinfo {author} {\bibfnamefont {J.~P.}\ \bibnamefont {Santos~Pires}},\ and\ \bibinfo {author} {\bibfnamefont {J.~M. V.~P.}\ \bibnamefont {Lopes}},\ }\bibfield  {title} {\bibinfo {title} {Unambiguous simulation of diffusive charge transport in disordered nanoribbons},\ }\href {https://doi.org/10.21468/SciPostPhys.17.6.149} {\bibfield  {journal} {\bibinfo  {journal} {SciPost Physics}\ }\textbf {\bibinfo {volume} {17}},\ \bibinfo {pages} {149} (\bibinfo {year} {2024})}\BibitemShut {NoStop}%
\bibitem [{\citenamefont {Santos~Pires}\ \emph {et~al.}(2020)\citenamefont {Santos~Pires}, \citenamefont {Amorim},\ and\ \citenamefont {Viana Parente~Lopes}}]{santos2020landauer}%
  \BibitemOpen
  \bibfield  {author} {\bibinfo {author} {\bibfnamefont {J.~P.}\ \bibnamefont {Santos~Pires}}, \bibinfo {author} {\bibfnamefont {B.}~\bibnamefont {Amorim}},\ and\ \bibinfo {author} {\bibfnamefont {J.~M.}\ \bibnamefont {Viana Parente~Lopes}},\ }\bibfield  {title} {\bibinfo {title} {Landauer transport as a quasisteady state on finite chains under unitary quantum dynamics},\ }\href {https://doi.org/10.1103/PhysRevB.101.104203} {\bibfield  {journal} {\bibinfo  {journal} {Physical Review B}\ }\textbf {\bibinfo {volume} {101}},\ \bibinfo {pages} {104203} (\bibinfo {year} {2020})}\BibitemShut {NoStop}%
\bibitem [{\citenamefont {Di~Ventra}(2008)}]{diventra2008electrical}%
  \BibitemOpen
  \bibfield  {author} {\bibinfo {author} {\bibfnamefont {M.}~\bibnamefont {Di~Ventra}},\ }\href@noop {} {\emph {\bibinfo {title} {Electrical transport in nanoscale systems}}}\ (\bibinfo  {publisher} {Cambridge University Press},\ \bibinfo {year} {2008})\BibitemShut {NoStop}%
\bibitem [{sup()}]{suppmaterial}%
  \BibitemOpen
  \href@noop {} {}\bibinfo {note} {See Supplemental Material at [URL will be inserted by publisher] for a detailed description of the Kernel Polynomial Method, derive the time-dependent Kubo formula for the conductivity, and compare the Kubo-Bastin and time-dependent formalisms. We also present the density of states for the polycrystalline and amorphous samples, explain the calculation of the momentum relaxation time, and show the orbital Hall effect at higher chemical potentials and over longer timescales for the amorphous sample. Finally, we describe the generation and characterization of the polycrystalline and amorphous samples and present their local density of states.}\BibitemShut {Stop}%
\bibitem [{\citenamefont {Mart\'{\i}nez~Romeral}\ \emph {et~al.}(2026)\citenamefont {Mart\'{\i}nez~Romeral}, \citenamefont {Canonico}, \citenamefont {Cummings},\ and\ \citenamefont {Roche}}]{Romeral2026LightHall}%
  \BibitemOpen
  \bibfield  {author} {\bibinfo {author} {\bibfnamefont {J.}~\bibnamefont {Mart\'{\i}nez~Romeral}}, \bibinfo {author} {\bibfnamefont {L.~M.}\ \bibnamefont {Canonico}}, \bibinfo {author} {\bibfnamefont {A.~W.}\ \bibnamefont {Cummings}},\ and\ \bibinfo {author} {\bibfnamefont {S.}~\bibnamefont {Roche}},\ }\bibfield  {title} {\bibinfo {title} {Real-space approach to light-induced hall transport in disordered materials},\ }\href {https://doi.org/10.1103/xs9m-1l27} {\bibfield  {journal} {\bibinfo  {journal} {Phys. Rev. B}\ }\textbf {\bibinfo {volume} {113}},\ \bibinfo {pages} {245405} (\bibinfo {year} {2026})}\BibitemShut {NoStop}%
\bibitem [{\citenamefont {Streda}(1982)}]{Streda1982Theory}%
  \BibitemOpen
  \bibfield  {author} {\bibinfo {author} {\bibfnamefont {P.}~\bibnamefont {Streda}},\ }\bibfield  {title} {\bibinfo {title} {Theory of quantised hall conductivity in two dimensions},\ }\href {https://doi.org/10.1088/0022-3719/15/22/005} {\bibfield  {journal} {\bibinfo  {journal} {Journal of Physics C: Solid State Physics}\ }\textbf {\bibinfo {volume} {15}},\ \bibinfo {pages} {L717} (\bibinfo {year} {1982})}\BibitemShut {NoStop}%
\bibitem [{\citenamefont {Bastin}\ \emph {et~al.}(1971)\citenamefont {Bastin}, \citenamefont {Lewiner}, \citenamefont {Betbeder-matibet},\ and\ \citenamefont {Nozieres}}]{Bastin1971Quantum}%
  \BibitemOpen
  \bibfield  {author} {\bibinfo {author} {\bibfnamefont {A.}~\bibnamefont {Bastin}}, \bibinfo {author} {\bibfnamefont {C.}~\bibnamefont {Lewiner}}, \bibinfo {author} {\bibfnamefont {O.}~\bibnamefont {Betbeder-matibet}},\ and\ \bibinfo {author} {\bibfnamefont {P.}~\bibnamefont {Nozieres}},\ }\bibfield  {title} {\bibinfo {title} {Quantum oscillations of the hall effect of a fermion gas with random impurity scattering},\ }\href {https://doi.org/10.1016/S0022-3697(71)80147-6} {\bibfield  {journal} {\bibinfo  {journal} {Journal of Physics and Chemistry of Solids}\ }\textbf {\bibinfo {volume} {32}},\ \bibinfo {pages} {1811–1824} (\bibinfo {year} {1971})}\BibitemShut {NoStop}%
\bibitem [{\citenamefont {Kubo}(1957)}]{Kubo1957Stadistical}%
  \BibitemOpen
  \bibfield  {author} {\bibinfo {author} {\bibfnamefont {R.}~\bibnamefont {Kubo}},\ }\bibfield  {title} {\bibinfo {title} {Statistical-mechanical theory of irreversible processes i. general theory and simple applications to magnetic and conduction problems},\ }\href {https://doi.org/10.1143/JPSJ.12.570} {\bibfield  {journal} {\bibinfo  {journal} {Journal of the Physical Society of Japan}\ }\textbf {\bibinfo {volume} {12}},\ \bibinfo {pages} {570–586} (\bibinfo {year} {1957})}\BibitemShut {NoStop}%
\bibitem [{\citenamefont {Bruus}\ and\ \citenamefont {Flensberg}(2004)}]{Bruus2004ManyBody}%
  \BibitemOpen
  \bibfield  {author} {\bibinfo {author} {\bibfnamefont {H.}~\bibnamefont {Bruus}}\ and\ \bibinfo {author} {\bibfnamefont {K.}~\bibnamefont {Flensberg}},\ }\bibfield  {title} {\bibinfo {title} {Many-body quantum theory in condensed matter physics: An introduction},\ }\bibfield  {journal} {\bibinfo  {journal} {Oxford University Press}\ }\href {https://doi.org/10.1093/oso/9780198566335.001.0001} {10.1093/oso/9780198566335.001.0001} (\bibinfo {year} {2004})\BibitemShut {NoStop}%
\bibitem [{\citenamefont {Reichl}(2016)}]{Reichl2016Modern}%
  \BibitemOpen
  \bibfield  {author} {\bibinfo {author} {\bibfnamefont {L.~E.}\ \bibnamefont {Reichl}},\ }\href@noop {} {\emph {\bibinfo {title} {A Modern Course in Statistical Physics}}}\ (\bibinfo  {publisher} {John Wiley \& Sons},\ \bibinfo {year} {2016})\BibitemShut {NoStop}%
\bibitem [{\citenamefont {Panhans}\ and\ \citenamefont {Ortmann}(2021)}]{PanhansOrtmann2021}%
  \BibitemOpen
  \bibfield  {author} {\bibinfo {author} {\bibfnamefont {M.}~\bibnamefont {Panhans}}\ and\ \bibinfo {author} {\bibfnamefont {F.}~\bibnamefont {Ortmann}},\ }\bibfield  {title} {\bibinfo {title} {Efficient time-domain approach for linear response functions},\ }\href {https://doi.org/10.1103/PhysRevLett.127.016601} {\bibfield  {journal} {\bibinfo  {journal} {Phys. Rev. Lett.}\ }\textbf {\bibinfo {volume} {127}},\ \bibinfo {pages} {016601} (\bibinfo {year} {2021})}\BibitemShut {NoStop}%
\bibitem [{\citenamefont {Fan}\ \emph {et~al.}(2021)\citenamefont {Fan}, \citenamefont {Garcia}, \citenamefont {Cummings}, \citenamefont {Barrios-Vargas}, \citenamefont {Panhans}, \citenamefont {Harju}, \citenamefont {Ortmann},\ and\ \citenamefont {Roche}}]{fan2021linear}%
  \BibitemOpen
  \bibfield  {author} {\bibinfo {author} {\bibfnamefont {Z.}~\bibnamefont {Fan}}, \bibinfo {author} {\bibfnamefont {J.~H.}\ \bibnamefont {Garcia}}, \bibinfo {author} {\bibfnamefont {A.~W.}\ \bibnamefont {Cummings}}, \bibinfo {author} {\bibfnamefont {J.~E.}\ \bibnamefont {Barrios-Vargas}}, \bibinfo {author} {\bibfnamefont {M.}~\bibnamefont {Panhans}}, \bibinfo {author} {\bibfnamefont {A.}~\bibnamefont {Harju}}, \bibinfo {author} {\bibfnamefont {F.}~\bibnamefont {Ortmann}},\ and\ \bibinfo {author} {\bibfnamefont {S.}~\bibnamefont {Roche}},\ }\bibfield  {title} {\bibinfo {title} {Linear scaling quantum transport methodologies},\ }\href {https://doi.org/10.1016/j.physrep.2020.12.001} {\bibfield  {journal} {\bibinfo  {journal} {Physics Reports}\ }\bibinfo {series} {Linear scaling quantum transport methodologies},\ \textbf {\bibinfo {volume} {903}},\ \bibinfo {pages} {1–69} (\bibinfo {year} {2021})}\BibitemShut {NoStop}%
\bibitem [{\citenamefont {Thonhauser}\ \emph {et~al.}(2005)\citenamefont {Thonhauser}, \citenamefont {Ceresoli}, \citenamefont {Vanderbilt},\ and\ \citenamefont {Resta}}]{Tonhauser2005}%
  \BibitemOpen
  \bibfield  {author} {\bibinfo {author} {\bibfnamefont {T.}~\bibnamefont {Thonhauser}}, \bibinfo {author} {\bibfnamefont {D.}~\bibnamefont {Ceresoli}}, \bibinfo {author} {\bibfnamefont {D.}~\bibnamefont {Vanderbilt}},\ and\ \bibinfo {author} {\bibfnamefont {R.}~\bibnamefont {Resta}},\ }\bibfield  {title} {\bibinfo {title} {Orbital magnetization in periodic insulators},\ }\href {https://doi.org/10.1103/PhysRevLett.95.137205} {\bibfield  {journal} {\bibinfo  {journal} {Phys. Rev. Lett.}\ }\textbf {\bibinfo {volume} {95}},\ \bibinfo {pages} {137205} (\bibinfo {year} {2005})}\BibitemShut {NoStop}%
\bibitem [{\citenamefont {Bhowal}\ and\ \citenamefont {Vignale}(2021)}]{Bhowal2021}%
  \BibitemOpen
  \bibfield  {author} {\bibinfo {author} {\bibfnamefont {S.}~\bibnamefont {Bhowal}}\ and\ \bibinfo {author} {\bibfnamefont {G.}~\bibnamefont {Vignale}},\ }\bibfield  {title} {\bibinfo {title} {Orbital hall effect as an alternative to valley hall effect in gapped graphene},\ }\href {https://doi.org/10.1103/PhysRevB.103.195309} {\bibfield  {journal} {\bibinfo  {journal} {Phys. Rev. B}\ }\textbf {\bibinfo {volume} {103}},\ \bibinfo {pages} {195309} (\bibinfo {year} {2021})}\BibitemShut {NoStop}%
\bibitem [{\citenamefont {Weiße}\ \emph {et~al.}(2006)\citenamefont {Weiße}, \citenamefont {Wellein}, \citenamefont {Alvermann},\ and\ \citenamefont {Fehske}}]{Weisse2006Kernel}%
  \BibitemOpen
  \bibfield  {author} {\bibinfo {author} {\bibfnamefont {A.}~\bibnamefont {Weiße}}, \bibinfo {author} {\bibfnamefont {G.}~\bibnamefont {Wellein}}, \bibinfo {author} {\bibfnamefont {A.}~\bibnamefont {Alvermann}},\ and\ \bibinfo {author} {\bibfnamefont {H.}~\bibnamefont {Fehske}},\ }\bibfield  {title} {\bibinfo {title} {The kernel polynomial method},\ }\href {https://doi.org/10.1103/revmodphys.78.275} {\bibfield  {journal} {\bibinfo  {journal} {Reviews of Modern Physics}\ }\textbf {\bibinfo {volume} {78}},\ \bibinfo {pages} {275–306} (\bibinfo {year} {2006})}\BibitemShut {NoStop}%
\bibitem [{\citenamefont {Foa~Torres}\ \emph {et~al.}(2014)\citenamefont {Foa~Torres}, \citenamefont {Roche},\ and\ \citenamefont {Charlier}}]{FoaTorresRocheChalier2014}%
  \BibitemOpen
  \bibfield  {author} {\bibinfo {author} {\bibfnamefont {L.~E.~F.}\ \bibnamefont {Foa~Torres}}, \bibinfo {author} {\bibfnamefont {S.}~\bibnamefont {Roche}},\ and\ \bibinfo {author} {\bibfnamefont {J.-C.}\ \bibnamefont {Charlier}},\ }\href@noop {} {\emph {\bibinfo {title} {Introduction to Graphene-Based Nanomaterials}}}\ (\bibinfo  {publisher} {Cambridge University Press},\ \bibinfo {address} {Cambridge, UK},\ \bibinfo {year} {2014})\BibitemShut {NoStop}%
\bibitem [{\citenamefont {Anderson}(1958)}]{Anderson1958}%
  \BibitemOpen
  \bibfield  {author} {\bibinfo {author} {\bibfnamefont {P.~W.}\ \bibnamefont {Anderson}},\ }\bibfield  {title} {\bibinfo {title} {Absence of diffusion in certain random lattices},\ }\href {https://doi.org/10.1103/PhysRev.109.1492} {\bibfield  {journal} {\bibinfo  {journal} {Phys. Rev.}\ }\textbf {\bibinfo {volume} {109}},\ \bibinfo {pages} {1492} (\bibinfo {year} {1958})}\BibitemShut {NoStop}%
\bibitem [{\citenamefont {Cummings}\ \emph {et~al.}(2019)\citenamefont {Cummings}, \citenamefont {Dubois}, \citenamefont {Charlier},\ and\ \citenamefont {Roche}}]{cummings2019universal}%
  \BibitemOpen
  \bibfield  {author} {\bibinfo {author} {\bibfnamefont {A.~W.}\ \bibnamefont {Cummings}}, \bibinfo {author} {\bibfnamefont {S.~M.-M.}\ \bibnamefont {Dubois}}, \bibinfo {author} {\bibfnamefont {J.-C.}\ \bibnamefont {Charlier}},\ and\ \bibinfo {author} {\bibfnamefont {S.}~\bibnamefont {Roche}},\ }\bibfield  {title} {\bibinfo {title} {Universal spin diffusion length in polycrystalline graphene},\ }\href {https://doi.org/10.1021/acs.nanolett.9b03112} {\bibfield  {journal} {\bibinfo  {journal} {Nano Letters}\ }\textbf {\bibinfo {volume} {19}},\ \bibinfo {pages} {7418} (\bibinfo {year} {2019})},\ \bibinfo {note} {pMID: 31532994},\ \Eprint {https://arxiv.org/abs/https://doi.org/10.1021/acs.nanolett.9b03112} {https://doi.org/10.1021/acs.nanolett.9b03112} \BibitemShut {NoStop}%
\bibitem [{\citenamefont {Kotakoski}\ and\ \citenamefont {Meyer}(2012)}]{Kotakoski2012Mechanical}%
  \BibitemOpen
  \bibfield  {author} {\bibinfo {author} {\bibfnamefont {J.}~\bibnamefont {Kotakoski}}\ and\ \bibinfo {author} {\bibfnamefont {J.~C.}\ \bibnamefont {Meyer}},\ }\bibfield  {title} {\bibinfo {title} {Mechanical properties of polycrystalline graphene based on a realistic atomistic model},\ }\href {https://doi.org/10.1103/PhysRevB.85.195447} {\bibfield  {journal} {\bibinfo  {journal} {Physical Review B}\ }\textbf {\bibinfo {volume} {85}},\ \bibinfo {pages} {195447} (\bibinfo {year} {2012})}\BibitemShut {NoStop}%
\bibitem [{\citenamefont {Lindsay}\ and\ \citenamefont {Broido}(2010)}]{Lindsay2010Optimized}%
  \BibitemOpen
  \bibfield  {author} {\bibinfo {author} {\bibfnamefont {L.}~\bibnamefont {Lindsay}}\ and\ \bibinfo {author} {\bibfnamefont {D.~A.}\ \bibnamefont {Broido}},\ }\bibfield  {title} {\bibinfo {title} {Optimized tersoff and brenner empirical potential parameters for lattice dynamics and phonon thermal transport in carbon nanotubes and graphene},\ }\href {https://doi.org/10.1103/PhysRevB.81.205441} {\bibfield  {journal} {\bibinfo  {journal} {Phys. Rev. B}\ }\textbf {\bibinfo {volume} {81}},\ \bibinfo {pages} {205441} (\bibinfo {year} {2010})}\BibitemShut {NoStop}%
\bibitem [{\citenamefont {Guerrero}\ \emph {et~al.}(2025)\citenamefont {Guerrero}, \citenamefont {Nguyen}, \citenamefont {Romeral}, \citenamefont {Cummings}, \citenamefont {Garcia}, \citenamefont {Charlier},\ and\ \citenamefont {Roche}}]{Guerrero2025}%
  \BibitemOpen
  \bibfield  {author} {\bibinfo {author} {\bibfnamefont {P.~A.}\ \bibnamefont {Guerrero}}, \bibinfo {author} {\bibfnamefont {V.-H.}\ \bibnamefont {Nguyen}}, \bibinfo {author} {\bibfnamefont {J.~M.}\ \bibnamefont {Romeral}}, \bibinfo {author} {\bibfnamefont {A.~W.}\ \bibnamefont {Cummings}}, \bibinfo {author} {\bibfnamefont {J.-H.}\ \bibnamefont {Garcia}}, \bibinfo {author} {\bibfnamefont {J.-C.}\ \bibnamefont {Charlier}},\ and\ \bibinfo {author} {\bibfnamefont {S.}~\bibnamefont {Roche}},\ }\bibfield  {title} {\bibinfo {title} {Disorder-induced delocalization in magic-angle twisted bilayer graphene},\ }\href {https://doi.org/10.1103/PhysRevLett.134.126301} {\bibfield  {journal} {\bibinfo  {journal} {Phys. Rev. Lett.}\ }\textbf {\bibinfo {volume} {134}},\ \bibinfo {pages} {126301} (\bibinfo {year} {2025})}\BibitemShut {NoStop}%
\bibitem [{\citenamefont {Sinova}\ \emph {et~al.}(2015)\citenamefont {Sinova}, \citenamefont {Valenzuela}, \citenamefont {Wunderlich}, \citenamefont {Back},\ and\ \citenamefont {Jungwirth}}]{Sinova2015}%
  \BibitemOpen
  \bibfield  {author} {\bibinfo {author} {\bibfnamefont {J.}~\bibnamefont {Sinova}}, \bibinfo {author} {\bibfnamefont {S.~O.}\ \bibnamefont {Valenzuela}}, \bibinfo {author} {\bibfnamefont {J.}~\bibnamefont {Wunderlich}}, \bibinfo {author} {\bibfnamefont {C.~H.}\ \bibnamefont {Back}},\ and\ \bibinfo {author} {\bibfnamefont {T.}~\bibnamefont {Jungwirth}},\ }\bibfield  {title} {\bibinfo {title} {Spin hall effects},\ }\href {https://doi.org/10.1103/RevModPhys.87.1213} {\bibfield  {journal} {\bibinfo  {journal} {Rev. Mod. Phys.}\ }\textbf {\bibinfo {volume} {87}},\ \bibinfo {pages} {1213} (\bibinfo {year} {2015})}\BibitemShut {NoStop}%
\bibitem [{\citenamefont {Zhang}\ \emph {et~al.}(2013)\citenamefont {Zhang}, \citenamefont {Zhang},\ and\ \citenamefont {Zhou}}]{Zhang2013RevCVD}%
  \BibitemOpen
  \bibfield  {author} {\bibinfo {author} {\bibfnamefont {Y.}~\bibnamefont {Zhang}}, \bibinfo {author} {\bibfnamefont {L.}~\bibnamefont {Zhang}},\ and\ \bibinfo {author} {\bibfnamefont {C.}~\bibnamefont {Zhou}},\ }\bibfield  {title} {\bibinfo {title} {Review of chemical vapor deposition of graphene and related applications},\ }\href {https://doi.org/10.1021/ar300203n} {\bibfield  {journal} {\bibinfo  {journal} {Accounts of Chemical Research}\ }\textbf {\bibinfo {volume} {46}},\ \bibinfo {pages} {2329} (\bibinfo {year} {2013})},\ \bibinfo {note} {pMID: 23480816},\ \Eprint {https://arxiv.org/abs/https://doi.org/10.1021/ar300203n} {https://doi.org/10.1021/ar300203n} \BibitemShut {NoStop}%
\bibitem [{\citenamefont {Choi}\ \emph {et~al.}(2022)\citenamefont {Choi}, \citenamefont {Yun}, \citenamefont {Won}, \citenamefont {Oh}, \citenamefont {Kim}, \citenamefont {Kim},\ and\ \citenamefont {Lee}}]{ChoiLarge2022}%
  \BibitemOpen
  \bibfield  {author} {\bibinfo {author} {\bibfnamefont {S.~H.}\ \bibnamefont {Choi}}, \bibinfo {author} {\bibfnamefont {S.~J.}\ \bibnamefont {Yun}}, \bibinfo {author} {\bibfnamefont {Y.~S.}\ \bibnamefont {Won}}, \bibinfo {author} {\bibfnamefont {C.~S.}\ \bibnamefont {Oh}}, \bibinfo {author} {\bibfnamefont {S.~M.}\ \bibnamefont {Kim}}, \bibinfo {author} {\bibfnamefont {K.~K.}\ \bibnamefont {Kim}},\ and\ \bibinfo {author} {\bibfnamefont {Y.~H.}\ \bibnamefont {Lee}},\ }\bibfield  {title} {\bibinfo {title} {Large-scale synthesis of graphene and other 2d materials towards industrialization},\ }\bibfield  {journal} {\bibinfo  {journal} {Nature Communications}\ }\textbf {\bibinfo {volume} {13}},\ \href {https://doi.org/10.1038/s41467-022-29182-y} {10.1038/s41467-022-29182-y} (\bibinfo {year} {2022})\BibitemShut {NoStop}%
\bibitem [{\citenamefont {Kotakoski}\ \emph {et~al.}(2011)\citenamefont {Kotakoski}, \citenamefont {Krasheninnikov}, \citenamefont {Kaiser},\ and\ \citenamefont {Meyer}}]{Kotakoski2011}%
  \BibitemOpen
  \bibfield  {author} {\bibinfo {author} {\bibfnamefont {J.}~\bibnamefont {Kotakoski}}, \bibinfo {author} {\bibfnamefont {A.~V.}\ \bibnamefont {Krasheninnikov}}, \bibinfo {author} {\bibfnamefont {U.}~\bibnamefont {Kaiser}},\ and\ \bibinfo {author} {\bibfnamefont {J.~C.}\ \bibnamefont {Meyer}},\ }\bibfield  {title} {\bibinfo {title} {From point defects in graphene to two-dimensional amorphous carbon},\ }\href {https://doi.org/10.1103/PhysRevLett.106.105505} {\bibfield  {journal} {\bibinfo  {journal} {Phys. Rev. Lett.}\ }\textbf {\bibinfo {volume} {106}},\ \bibinfo {pages} {105505} (\bibinfo {year} {2011})}\BibitemShut {NoStop}%
\bibitem [{\citenamefont {Van~Tuan}\ \emph {et~al.}(2012)\citenamefont {Van~Tuan}, \citenamefont {Kumar}, \citenamefont {Roche}, \citenamefont {Ortmann}, \citenamefont {Thorpe},\ and\ \citenamefont {Ordejon}}]{VanTuan2012}%
  \BibitemOpen
  \bibfield  {author} {\bibinfo {author} {\bibfnamefont {D.}~\bibnamefont {Van~Tuan}}, \bibinfo {author} {\bibfnamefont {A.}~\bibnamefont {Kumar}}, \bibinfo {author} {\bibfnamefont {S.}~\bibnamefont {Roche}}, \bibinfo {author} {\bibfnamefont {F.}~\bibnamefont {Ortmann}}, \bibinfo {author} {\bibfnamefont {M.~F.}\ \bibnamefont {Thorpe}},\ and\ \bibinfo {author} {\bibfnamefont {P.}~\bibnamefont {Ordejon}},\ }\bibfield  {title} {\bibinfo {title} {Insulating behavior of an amorphous graphene membrane},\ }\href {https://doi.org/10.1103/PhysRevB.86.121408} {\bibfield  {journal} {\bibinfo  {journal} {Phys. Rev. B}\ }\textbf {\bibinfo {volume} {86}},\ \bibinfo {pages} {121408} (\bibinfo {year} {2012})}\BibitemShut {NoStop}%
\bibitem [{\citenamefont {Toh}\ \emph {et~al.}(2026)\citenamefont {Toh}, \citenamefont {Grebenko}, \citenamefont {Karadeniz}, \citenamefont {Bhat}, \citenamefont {He}, \citenamefont {Zhang}, \citenamefont {Shi}, \citenamefont {Rogov}, \citenamefont {Noskova}, \citenamefont {Starkov}, \citenamefont {Alekseeva}, \citenamefont {Iakoubovskii}, \citenamefont {Tee}, \citenamefont {Vyalikh}, \citenamefont {Makarova}, \citenamefont {Fedorov}, \citenamefont {Sassi}, \citenamefont {Bosman}, \citenamefont {Kamiuchi}, \citenamefont {Sato}, \citenamefont {Suenaga},\ and\ \citenamefont {{\"O}zyilmaz}}]{Toh2026Atomically}%
  \BibitemOpen
  \bibfield  {author} {\bibinfo {author} {\bibfnamefont {C.-T.}\ \bibnamefont {Toh}}, \bibinfo {author} {\bibfnamefont {A.~K.}\ \bibnamefont {Grebenko}}, \bibinfo {author} {\bibfnamefont {U.}~\bibnamefont {Karadeniz}}, \bibinfo {author} {\bibfnamefont {U.}~\bibnamefont {Bhat}}, \bibinfo {author} {\bibfnamefont {Y.}~\bibnamefont {He}}, \bibinfo {author} {\bibfnamefont {H.}~\bibnamefont {Zhang}}, \bibinfo {author} {\bibfnamefont {L.}~\bibnamefont {Shi}}, \bibinfo {author} {\bibfnamefont {I.}~\bibnamefont {Rogov}}, \bibinfo {author} {\bibfnamefont {D.}~\bibnamefont {Noskova}}, \bibinfo {author} {\bibfnamefont {A.}~\bibnamefont {Starkov}}, \bibinfo {author} {\bibfnamefont {A.~A.}\ \bibnamefont {Alekseeva}}, \bibinfo {author} {\bibfnamefont {K.~V.}\ \bibnamefont {Iakoubovskii}}, \bibinfo {author} {\bibfnamefont {C.~C.}\ \bibnamefont {Tee}}, \bibinfo {author} {\bibfnamefont {D.~V.}\ \bibnamefont {Vyalikh}}, \bibinfo {author} {\bibfnamefont {A.}~\bibnamefont {Makarova}}, \bibinfo {author} {\bibfnamefont {A.}~\bibnamefont {Fedorov}}, \bibinfo {author} {\bibfnamefont {L.~M.}\ \bibnamefont {Sassi}}, \bibinfo {author} {\bibfnamefont {M.}~\bibnamefont {Bosman}}, \bibinfo {author} {\bibfnamefont {N.}~\bibnamefont {Kamiuchi}}, \bibinfo {author} {\bibfnamefont {Y.}~\bibnamefont {Sato}}, \bibinfo {author} {\bibfnamefont {K.}~\bibnamefont {Suenaga}},\ and\ \bibinfo {author} {\bibfnamefont {B.}~\bibnamefont {{\"O}zyilmaz}},\ }\bibfield  {title} {\bibinfo {title} {Atomically thin amorphous carbon with an ultralow dielectric constant},\ }\href {https://doi.org/10.1038/s41928-026-01685-2} {\bibfield  {journal} {\bibinfo  {journal} {Nature Electronics}\ }\textbf {\bibinfo {volume} {9}},\ \bibinfo {pages} {1018} (\bibinfo {year} {2026})}\BibitemShut {NoStop}%
\bibitem [{\citenamefont {SK}\ \emph {et~al.}(2026)\citenamefont {SK}, \citenamefont {Zhang}, \citenamefont {Grebenko}, \citenamefont {Herasymchuk}, \citenamefont {Shivajirao}, \citenamefont {Zhang}, \citenamefont {Nelson}, \citenamefont {Tong}, \citenamefont {Singh}, \citenamefont {Kimiuchi}, \citenamefont {Sato}, \citenamefont {Suenaga}, \citenamefont {Toh}, \citenamefont {Romer}, \citenamefont {Adam}, \citenamefont {Yazyev}, \citenamefont {Ozyilmaz},\ and\ \citenamefont {Weber}}]{SK2026Quantum}%
  \BibitemOpen
  \bibfield  {author} {\bibinfo {author} {\bibfnamefont {R.}~\bibnamefont {SK}}, \bibinfo {author} {\bibfnamefont {H.}~\bibnamefont {Zhang}}, \bibinfo {author} {\bibfnamefont {A.~K.}\ \bibnamefont {Grebenko}}, \bibinfo {author} {\bibfnamefont {A.}~\bibnamefont {Herasymchuk}}, \bibinfo {author} {\bibfnamefont {R.}~\bibnamefont {Shivajirao}}, \bibinfo {author} {\bibfnamefont {H.}~\bibnamefont {Zhang}}, \bibinfo {author} {\bibfnamefont {A.}~\bibnamefont {Nelson}}, \bibinfo {author} {\bibfnamefont {Z.~J.}\ \bibnamefont {Tong}}, \bibinfo {author} {\bibfnamefont {G.}~\bibnamefont {Singh}}, \bibinfo {author} {\bibfnamefont {N.}~\bibnamefont {Kimiuchi}}, \bibinfo {author} {\bibfnamefont {Y.}~\bibnamefont {Sato}}, \bibinfo {author} {\bibfnamefont {K.}~\bibnamefont {Suenaga}}, \bibinfo {author} {\bibfnamefont {C.~T.}\ \bibnamefont {Toh}}, \bibinfo {author} {\bibfnamefont {R.~A.}\ \bibnamefont {Romer}}, \bibinfo {author} {\bibfnamefont {S.}~\bibnamefont {Adam}}, \bibinfo {author} {\bibfnamefont {O.~V.}\ \bibnamefont {Yazyev}}, \bibinfo {author} {\bibfnamefont {B.}~\bibnamefont {Ozyilmaz}},\ and\ \bibinfo {author} {\bibfnamefont {B.}~\bibnamefont {Weber}},\ }\bibfield  {title} {\bibinfo {title} {Quantum criticality in monolayer amorphous carbon},\ }\bibfield  {journal} {\bibinfo  {journal} {ArXiV}\ }\href {https://doi.org/10.48550/arXiv.2605.14349} {10.48550/arXiv.2605.14349} (\bibinfo {year} {2026}),\ \bibinfo {note} {arXiv:2605.14349 [cond-mat.dis-nn]}\BibitemShut {NoStop}%
\bibitem [{\citenamefont {Liu}\ and\ \citenamefont {Culcer}(2024)}]{liu2024dominance}%
  \BibitemOpen
  \bibfield  {author} {\bibinfo {author} {\bibfnamefont {H.}~\bibnamefont {Liu}}\ and\ \bibinfo {author} {\bibfnamefont {D.}~\bibnamefont {Culcer}},\ }\bibfield  {title} {\bibinfo {title} {Dominance of extrinsic scattering mechanisms in the orbital hall effect: Graphene, transition metal dichalcogenides, and topological antiferromagnets},\ }\href {https://doi.org/10.1103/PhysRevLett.132.186302} {\bibfield  {journal} {\bibinfo  {journal} {Physical Review Letters}\ }\textbf {\bibinfo {volume} {132}},\ \bibinfo {pages} {186302} (\bibinfo {year} {2024})}\BibitemShut {NoStop}%
\bibitem [{\citenamefont {Salvador-S\'anchez}\ \emph {et~al.}(2024)\citenamefont {Salvador-S\'anchez}, \citenamefont {Canonico}, \citenamefont {P\'erez-Rodr\'{\i}guez}, \citenamefont {Cysne}, \citenamefont {Baba}, \citenamefont {Cleric\`o}, \citenamefont {Vila}, \citenamefont {Vaquero}, \citenamefont {Delgado-Notario}, \citenamefont {Caridad}, \citenamefont {Watanabe}, \citenamefont {Taniguchi}, \citenamefont {Molina}, \citenamefont {Dom\'{\i}nguez-Adame}, \citenamefont {Roche}, \citenamefont {Diez}, \citenamefont {Rappoport},\ and\ \citenamefont {Amado}}]{SalvadorSanchez2024}%
  \BibitemOpen
  \bibfield  {author} {\bibinfo {author} {\bibfnamefont {J.}~\bibnamefont {Salvador-S\'anchez}}, \bibinfo {author} {\bibfnamefont {L.~M.}\ \bibnamefont {Canonico}}, \bibinfo {author} {\bibfnamefont {A.}~\bibnamefont {P\'erez-Rodr\'{\i}guez}}, \bibinfo {author} {\bibfnamefont {T.~P.}\ \bibnamefont {Cysne}}, \bibinfo {author} {\bibfnamefont {Y.}~\bibnamefont {Baba}}, \bibinfo {author} {\bibfnamefont {V.}~\bibnamefont {Cleric\`o}}, \bibinfo {author} {\bibfnamefont {M.}~\bibnamefont {Vila}}, \bibinfo {author} {\bibfnamefont {D.}~\bibnamefont {Vaquero}}, \bibinfo {author} {\bibfnamefont {J.~A.}\ \bibnamefont {Delgado-Notario}}, \bibinfo {author} {\bibfnamefont {J.~M.}\ \bibnamefont {Caridad}}, \bibinfo {author} {\bibfnamefont {K.}~\bibnamefont {Watanabe}}, \bibinfo {author} {\bibfnamefont {T.}~\bibnamefont {Taniguchi}}, \bibinfo {author} {\bibfnamefont {R.~A.}\ \bibnamefont {Molina}}, \bibinfo {author} {\bibfnamefont {F.}~\bibnamefont {Dom\'{\i}nguez-Adame}}, \bibinfo {author} {\bibfnamefont {S.}~\bibnamefont {Roche}}, \bibinfo {author} {\bibfnamefont {E.}~\bibnamefont {Diez}}, \bibinfo {author} {\bibfnamefont {T.~G.}\ \bibnamefont {Rappoport}},\ and\ \bibinfo {author} {\bibfnamefont {M.}~\bibnamefont {Amado}},\ }\bibfield  {title} {\bibinfo {title} {Generation and control of nonlocal chiral currents in graphene superlattices by orbital hall effect},\ }\href {https://doi.org/10.1103/PhysRevResearch.6.023212} {\bibfield  {journal} {\bibinfo  {journal} {Phys. Rev. Res.}\ }\textbf {\bibinfo {volume} {6}},\ \bibinfo {pages} {023212} (\bibinfo {year} {2024})}\BibitemShut {NoStop}%
\end{thebibliography}%

\end{document}